\documentclass[aps,prb,twocolumn,groupedaddress,showpacs,floatfix,altaffilletter,longbibliography, include graphics]{revtex4-2}
\usepackage{graphicx}
\usepackage{grffile}
\usepackage{amsmath}
\usepackage{amssymb}
\usepackage{bm}
\usepackage{color}
\usepackage[dvipsnames]{xcolor}
\usepackage{amsmath,amssymb,amsfonts}
\usepackage{epsfig}
\usepackage{times}
\usepackage[colorlinks,bookmarks=false,citecolor=blue,linkcolor=red,urlcolor=blue]{hyperref}
\usepackage{gensymb}
\usepackage[toc,page]{appendix}
\usepackage{float}

\newcommand{\fref}[1]{Fig.~\ref{#1}}
\newcommand{\eref}[1]{Eq.~(\ref{#1})} 
\newcommand{\sref}[1]{Sec.~(\ref{#1})}

\begin{document}

\title{The Two-Particle Self-Consistent Approach for Multiorbital models: application to the Emery model}

\author{C. Gauvin-Ndiaye$^{1}$, J. Leblanc$^{1}$, S. Marin$^1$, N. Martin$^1$, D. Lessnich$^{1,2}$ and A.-M.S.~Tremblay$^{1}$}
\affiliation{$^1$D{\'e}partement de Physique, Institut quantique, and RQMP Universit{\'e} de Sherbrooke, Sherbrooke, Qu{\'e}bec, Canada  J1K 2R1 }
\affiliation{$^2$Institute for Theoretical Physics, Goethe University Frankfurt, Max-von-Laue-Strasse 1, 60438 Frankfurt am Main, Germany}
\date{\today}
\begin{abstract}
The Emery model, or three-band Hubbard model, is a Hamiltonian that is thought to contain much of the physics of cuprate superconductors. This model includes two noninteracting $p$ orbitals and one interacting $d$ orbital per unit cell. Few methods that can solve multiorbital interacting Hamiltonians reliably and efficiently exist. Here, we introduce an application of the two particle self-consistent (TPSC) approach to the Emery model. We construct this method within the framework of the TPSC+DMFT method, which can be seen as a way to introduce nonlocal corrections to dynamical mean-field theory (DMFT). We show that interacting orbital densities, rather than the noninteracting ones, must be used in the calculations. For the Emery model, we find that at constant bare interaction $U$, the vertex for spin fluctuations, $U_{sp}$, decreases rapidly with filling. This may be one of the factors that contributes to electron-doped cuprates appearing less correlated than hole-doped ones. More generally, our work opens the road to the application of the TPSC approach to spin fluctuations in multiorbital models.
\end{abstract}

\maketitle
%\section{Remarques}
%(Fausse section juste pour mettre des commentaires)

% Introduction
\section{Introduction}

The one-band Hubbard model is one of the most widely used model hamiltonians for the study of strongly correlated systems. A typical example is the cuprate superconductors, which have been modeled extensively with the Hubbard model \cite{Scalapino_2012}. The one-band Hubbard model is still a challenging problem despite its apparent simplicity. The development of multiple approximate numerical approaches has been motivated by this model~\cite{LeBlanc_Antipov_Becca_Bulik_Chan_Chung_Deng_Ferrero:2015,Schafer_2021}. %These methods include the dynamical mean field approximation (DMFT) \textcolor{red}{Références}, and its cluster \textcolor{red}{Références} and diagrammatic \textcolor{red}{Références} extensions. In some regime of parameters, quantum Monte Carlo simulations offer exact results against which the approximate methods can be benchmarked \textcolor{red}{Références}.

However, multiorbital interactions that are beyond the scope of the Hubbard model need to be taken into account in order to capture the physics of many systems of interest. This is thought to be the case in unconventional superconductors such as the nickelates \cite{Held_2022}, strontium ruthenate \cite{Acharya_2017,Gingras_Tremblay:2022}, iron-based superconductors \cite{Yin_2011,Nourafkan_Kotliar_Tremblay_2016,Si_2023}, and heavy-fermion superconductors \cite{Hazra_2023}.

The Emery model, also called three-band Hubbard model or Emery-VSA (Varma, Schmitt-Rink, Abrahams) model, is a multiorbital Hamiltonian that was introduced for the study of superconductivity in the cuprates \cite{Emery_1987,Varma_1987}. Since then, numerous aspects of the model were studied by different numerical and theoretical techniques. For instance, superconductivity in the model was studied with multiple techniques \cite{Yanagisawa_2009,Weber_2012,Fratino_2016,Dash_2019,Zegrodnik_2019,Kowalski_2021,Zegrodnik_2021,Mai_2021}. Moreover, general studies of the phase diagram \cite{Kung_2016,Cui_2020}, antiferromagnetism \cite{Yanagisawa_2009}, pseudogap \cite{Fratino_2016,Dash_2019} and Mott insulator \cite{Go_2015} were realized. The Emery model enabled the study of electronic correlations in the cuprates \cite{Medici_2009,Weber_2010,Wang_2011} and of the asymmetry between the hole- and electron-doped cuprates \cite{Weber_2010,Ogura_2015}. Multiorbital hamiltonians such as the Emery model are even more difficult to solve numerically than the one-band Hubbard model. However, their physical relevance makes the development of reliable, computationally inexpensive methods of high interest. 

The two-particle self-consistent approach (TPSC) is a many-body method that was first developed for the one-band Hubbard model \cite{Vilk_1996,Vilk_1997,Tremblay_2011}. The TPSC approach for the one-band Hubbard model is a nonperturbative method that satisfies the Pauli principle, the Mermin-Wagner theorem, local charge and spin sum rules, and conservation laws. This method is valid in the weak to intermediate interaction regime of the Hubbard model. It was benchmarked extensively against exact quantum and diagrammatic Monte Carlo simulations \cite{Vilk_1994,Veilleux_1995,Vilk_1996,Vilk_1997,Moukouri_2000,Kyung_2001,Tremblay_2011,Schafer_2021,Martin_2023,Gauvin-Ndiaye_2023}. This approach is reliable in its regime of validity, and is numerically efficient. Because of these reasons, the method has been extended to multisite \cite{Aizawa_2015,Arya_2015,Zantout_2018,Pizarro_2020} and multiorbital \cite{Miyahara_2013,Ogura_2015,Zantout_2019,Zantout_2021} hamiltonians. Moreover, a recent improvement of TPSC for the one-band Hubbard model, the TPSC+DMFT approach \cite{Martin_2023}, has also been applied to multiorbital hamiltonians \cite{Zantout_2022}.

In this paper, we revisit the development of the multiorbital TPSC approach with a specific focus in mind: its application to the Emery model, which is relevant to the study of the cuprate superconductors. The Emery model is an ``extreme'' case in the sense that the two types of orbitals it describes are highly nondegenerate. Though the  extension of TPSC to the multiorbital model presented in Ref. \cite{Miyahara_2013} has been applied to the Emery model \cite{Ogura_2015}, here, we present a specific formulation of the method within the TPSC+DMFT framework. This problem highlights a specific difficulty in extending the TPSC approach to multiorbital models that was not considered in previous works. More specifically, we find that the densities that enter the calculation of the sum rules for the spin and charge susceptibilities must be calculated at the interacting level when dealing with hamiltonians where the orbitals are not degenerate. This aspect of the TPSC approach for multiorbital models has not been considered before and is important for further applications of the method to more generalized Hamiltonians, including Hund's interaction for example.

After introducing the formalism for a  generic Hamiltonian with density-density interactions in the following section, we apply the general formalism to the Emery model in section~\ref{sec:Emery}. The specific TPSC+DMFT approach to that model is described in section~\ref{sec:TPSC+DMFT}. While the derivations in these sections may seem daunting, the equations to be solved are rather simple and intuitive. The methodology with reference to the appropriate equations is summarized in section~\ref{sec:Methodology_TPSC+DMFT}. Finally, in section \ref{sec:results}, we apply the method to a specific realization of the Emery model to investigate the impacts of the choice of input orbital densities on the results. 

% Modèle
% Modèle
\section{Generic multi-orbital model}
We first introduce a generic hamiltonian with density-density interactions in \sref{sec:hamiltonian_nn}. We provide exact equations for its self-energy, susceptibilities and vertices in the source field framework developed by Schwinger and Martin \cite{Martin_1959, Kadanoff_1962} in \sref{sec:self_energy_nn} and \sref{sec:chis_gammas_nn}. We provide a complete, detailed calculation of all the results shown in this section in Appendix \ref{app:details_source_fields}.

\subsection{Hamiltonian with density-density interactions}
\label{sec:hamiltonian_nn}
We consider a simple multiorbital Hamiltonian of the form
%%%%%%%%%%%%%%%%
\begin{equation}
    H = \sum_{ij}\sum_\sigma\sum_{\alpha \beta}t_{ij,\alpha \beta}c^\dagger_{i\alpha\sigma}c_{j\beta \sigma} + \sum_{i\alpha\beta}U_{\alpha \beta}n_{i\alpha \uparrow}n_{i\beta\downarrow},
\end{equation}
%%%%%%%%%%%%%%%%
where $\alpha$ and $\beta$ are orbital indices, $i$ and $j$ are site indices, $t_{ij,\alpha\beta}$ is a hopping parameter from orbital $\beta$ on site $j$ to orbital $\alpha$ on site $i$, $c^{(\dagger)}_{i\alpha\sigma}$ is the annihilation (creation) operator for an electron of spin $\sigma$ on site $i$ and orbital $\alpha$, $n_{i\alpha\sigma}$ is the number operator for electrons of spin $\sigma$ on site $i$ and orbital $\alpha$, and $U_{\alpha\beta}$ is the on-site Hubbard interaction strength between orbitals $\alpha$ and $\beta$. As we will discuss later, the Emery model corresponds to the specific case with three orbitals denoted by the indices $d,~p_x$ and $p_y$, and where only the $d$ orbitals are correlated.

\subsection{Functional derivatives for the self-energy}
\label{sec:self_energy_nn}

In this section, we calculate the Green's function in the presence of a source field $\phi$. We introduce the partition function in the presence of a source field $\phi$
%%%%%%%%%%%%%%%%
\begin{equation}
    Z[\phi] = \left \langle T_\tau \exp\left[-c^\dagger_{\bar{\alpha}\bar{\sigma}}(\bar{1})\phi_{\bar{\alpha}\bar{\beta}\bar{\sigma}}(\bar{1},\bar{2})c_{\bar{\beta}\bar{\sigma}}(\bar{2})\right]\right \rangle,
\end{equation}
%%%%%%%%%%%%%%%%
from which we also define 
\begin{equation}
    S[\phi] = \exp\left[-c^\dagger_{\bar{\alpha}\bar{\sigma}}(\bar{1})\phi_{\bar{\alpha}\bar{\beta}\bar{\sigma}}(\bar{1},\bar{2})c_{\bar{\beta}\bar{\sigma}}(\bar{2})\right].
\end{equation}
%%%%%%%%%%%%%%%%
We use the notation $(\mathbf{r}_1,\tau_1) \equiv (1)$, where $\tau_1$ is an imaginary time. Moreover, the bars above symbols denote sums and integrals over all the degrees of freedom concerned: $\bar{\alpha} = \sum_\alpha$, $\bar{1} = \sum_{\mathbf{r}_1}\int_0^{\beta}d\tau_1$. Then, the Green's function in the presence of the source field $\phi$ is defined as
%%%%%%%%%%%%%%%%
\begin{align}
    \mathcal{G}_{\alpha\beta\sigma}(1,2)_\phi&=-\frac{\delta \ln{(Z[\phi])}}{\delta \phi_{\beta \alpha \sigma}(2,1)},\\
    &= -\left \langle T_\tau c_{\alpha \sigma}(1)c^\dagger_{\beta\sigma}(2)\right \rangle_\phi,
\end{align}
%%%%%%%%%%%%%%%%
where the average of an operator $O(\tau_1,\dots,\tau_N)$ in the presence of the source field is defined as
%%%%%%%%%%%%%%%%
\begin{equation}
    \left \langle T_\tau O(\tau_1,\dots,\tau_N)\right \rangle_\phi \equiv \frac{1}{Z[\phi]}\left \langle T_\tau O(\tau_1,\dots,\tau_N)S[\phi]\right \rangle.
\end{equation}
%%%%%%%%%%%%%%%%

We calculate the equations of motion for the Green's function by computing its imaginary time derivative $\frac{\partial \mathcal{G}_{\alpha\beta\sigma}(1,2)_\phi}{\partial \tau_1}$. As detailed in Appendix \ref{app:details_source_fields}, the self-energy is obtained from the commutator of the annihilation operator $c_{\alpha\sigma}(\tau_1)$ that appears in the Green's function with the interaction term of the Hamiltonian. From \eref{eq:app_selfEnergy_gen}, we hence define the self-energy as 
%%%%%%%%%%%%%%%%
\begin{align}
    \Sigma_{\alpha\bar{\gamma}\sigma}&(1,\bar{3})_\phi \mathcal{G}_{\bar{\gamma}\beta\sigma}(\bar{3},2)_\phi = \nonumber \\
    &-U_{\alpha\bar{\gamma}}\left \langle T_\tau c^\dagger_{\bar{\gamma}-\sigma}(1^+)c_{\bar{\gamma}-\sigma}(1)c_{\alpha\sigma}(1)c^\dagger_{\beta\sigma}(2)\right \rangle_\phi.
    \label{eq:selfEnergy_gen}
\end{align}
%%%%%%%%%%%%%%%%

This result is not unique: we can consider the same approach but perform the derivative with respect to $\tau_2$ instead of $\tau_1$. In that case, the self-energy is obtained from the commutator of the creation operator $c^\dagger_{\beta\sigma}(\tau_2)$ that appears in the Green's function with the interaction term of the hamiltonian. From \eref{eq:app_selfEnergy_gen_tau2}, the resulting second equation for the self-energy is
%%%%%%%%%%%%%%%%
\begin{align}
    \mathcal{G}_{\alpha \bar{\gamma}\sigma}&(1,\bar{3})_\phi \Sigma_{\bar{\gamma}\beta\sigma}(\bar{3},2)_\phi =\nonumber \\
    &-U_{\bar{\gamma}\beta}\left \langle T_\tau c^\dagger_{\bar{\gamma}-\sigma}(2^+)c_{\bar{\gamma}-\sigma}(2)c_{\alpha\sigma}(1)c^\dagger_{\beta\sigma}(2)\right \rangle_\phi.
    \label{eq:selfEnergy_gen_tau2}
\end{align}
%%%%%%%%%%%%%%%%

Both implicit expressions for the self-energy, \eref{eq:selfEnergy_gen} and \eref{eq:selfEnergy_gen_tau2}, should be satisfied simultaneously.

The interacting Green's function is calculated from the self-energy using the Dyson equation
%%%%%%%%%%%%%%%%
\begin{align}
    \mathcal{G}_{\alpha\beta\sigma}(1,2)_\phi^{-1} = &\mathcal{G}^{(0)}_{\alpha\beta\sigma}(1,2)^{-1}\nonumber \\
    &-\phi_{\alpha\beta\sigma}(1,2)-\Sigma_{\alpha\beta\sigma}(1,2)_\phi.
    \label{eq:Dyson_main_text}
\end{align}

\subsection{Susceptibilities and vertices}
\label{sec:chis_gammas_nn}
We now turn to the calculation of the susceptibilities. The calculations done in this section should be valid for a general multiorbital model. 

We first define generalized density and spin operators as
%%%%%%%%%%%%%%%%
\begin{equation}
    n_{\alpha\beta\sigma}(1,2) \equiv c^{\dagger}_{\beta\sigma}(2)c_{\alpha\sigma}(1),
\end{equation}
%%%%%%%%%%%%%%%%
%%%%%%%%%%%%%%%%
\begin{equation}
    n_{\alpha\beta}(1,2) \equiv n_{\alpha\beta\uparrow}(1,2) + n_{\alpha\beta\downarrow}(1,2),
\end{equation}
%%%%%%%%%%%%%%%%
%%%%%%%%%%%%%%%%
\begin{equation}
    S^z_{\alpha\beta}(1,2) \equiv n_{\alpha\beta\uparrow}(1,2) - n_{\alpha\beta\downarrow}(1,2).
\end{equation}
%%%%%%%%%%%%%%%%

Next, we define generalized susceptibilities as
%%%%%%%%%%%%%%%%
\begin{align}
    \chi^{+}_{\alpha\beta\gamma\delta}(1,2;3,4)_\phi & \equiv \left \langle T_\tau n_{\alpha\beta}(1,2)n_{\gamma\delta}(3,4)\right \rangle_\phi \nonumber \\
    &-\left \langle T_\tau n_{\alpha\beta}(1,2)\right \rangle_\phi \left \langle T_\tau n_{\gamma\delta}(3,4)\right \rangle_\phi,
\end{align}
%%%%%%%%%%%%%%%%
\begin{align}
    \chi^{-}_{\alpha\beta\gamma\delta}(1,2;3,4)_\phi &\equiv \left \langle T_\tau S^z_{\alpha\beta}(1,2)S^z_{\gamma\delta}(3,4)\right \rangle_\phi \nonumber \\
    &-\left \langle T_\tau S^z_{\alpha\beta}(1,2)\right \rangle_\phi \left \langle T_\tau S^z_{\gamma\delta}(3,4)\right \rangle_\phi.
\end{align}
%%%%%%%%%%%%%%%%

The non-interacting susceptibility can be written from the Green's functions as
%%%%%%%%%%%%%%%%
\begin{equation}
    \chi^{(0)}_{\alpha\beta\gamma\delta}(1,2;3,4)_\phi \equiv -2\mathcal{G}_{\alpha\delta\sigma}(1,4)_\phi\mathcal{G}_{\gamma\beta\sigma}(3,2)_\phi.
\end{equation}
%%%%%%%%%%%%%%%%
The generalized susceptibilities can also be written as functional derivatives of the Green's function
%%%%%%%%%%%%%%%%
\begin{align}
    \chi^{+}_{\alpha\beta\gamma\delta}(1,2;3,4)_\phi &= -\sum_{\sigma\sigma'}\frac{\delta \mathcal{G}_{\alpha\beta\sigma}(1,2)_\phi}{\delta \phi_{\delta\gamma\sigma'}(4,3)}, \label{eq:chiplus_derivs}\\
    \chi^{-}_{\alpha\beta\gamma\delta}(1,2;3,4)_\phi &= -\sum_{\sigma\sigma'}\sigma\sigma'\frac{\delta \mathcal{G}_{\alpha\beta\sigma}(1,2)_\phi}{\delta \phi_{\delta\gamma\sigma'}(4,3)}\label{eq:chiminus_derivs}.
\end{align}
%%%%%%%%%%%%%%%%
Using the Dyson equation \eqref{eq:Dyson_main_text}, the identity
%%%%%%%%%%%%%%%%
\begin{equation}
    \frac{\delta \mathcal{G}_{\alpha\beta\sigma}(1,2)_\phi}{\delta \phi_{\delta\gamma\sigma'}(4,3)} = -
    \mathcal{G}_{\alpha\bar{\eta}\sigma}(1,\bar{5})_\phi\frac{\delta \mathcal{G}_{\bar{\eta}\bar{\xi}\sigma}(\bar{5},\bar{6})_\phi^{-1}}{\delta \phi_{\delta\gamma\sigma'}(4,3)}\mathcal{G}_{\bar{\xi}\beta\sigma}(\bar{6},2)_\phi,
\end{equation}
%%%%%%%%%%%%%%%%
and recalling that the dependence of the self-energy on $\phi$ occurs only through the functional dependence of the Green's function on $\phi$, we can rewrite the generalized susceptibilities as follows
%%%%%%%%%%%%%%%%
\begin{align}
    \chi^{\pm}_{\alpha\beta\gamma\delta}&(1,2;3,4)_\phi = \chi^{(0)}_{\alpha\beta\gamma\delta}(1,2;3,4)_\phi \mp \nonumber \\
    &\frac{1}{2} \chi^{(0)}_{\alpha\beta\bar{\xi}\bar{\eta}}(1,2;\bar{6},\bar{5})_\phi\Gamma^{\pm}_{\bar{\xi}\bar{\eta}\bar{\mu}\bar{\nu}}(\bar{6},\bar{5},\bar{7},\bar{8})_\phi\chi^{\pm}_{\bar{\mu}\bar{\nu}\gamma\delta}(\bar{7},\bar{8};3,4)_\phi,
    \label{eq:gen_chipm}
\end{align}
%%%%%%%%%%%%%%%%
where the vertices $\Gamma^{\pm}_{\bar{\xi}\bar{\eta}\bar{\mu}\bar{\nu}}(\bar{6},\bar{5},\bar{7},\bar{8})_\phi$ are defined as
%%%%%%%%%%%%%%%%
\begin{equation}
    \Gamma^{\pm}_{\bar{\xi}\bar{\eta}\bar{\mu}\bar{\nu}}(\bar{6},\bar{5},\bar{7},\bar{8})_\phi = \frac{\delta \Sigma_{\bar{\eta}\bar{\xi}\downarrow}(\bar{5},\bar{6})_\phi}{\delta \mathcal{G}_{\bar{\mu}\bar{\nu}\uparrow}(\bar{7},\bar{8})_\phi} \pm \frac{\delta \Sigma_{\bar{\eta}\bar{\xi}\uparrow}(\bar{5},\bar{6})_\phi}{\delta \mathcal{G}_{\bar{\mu}\bar{\nu}\uparrow}(\bar{7},\bar{8})_\phi}.
\end{equation}
%%%%%%%%%%%%%%%%

% Application au modèle d'Emery
\begin{figure}
    \centering
    \includegraphics[width=0.8\columnwidth]{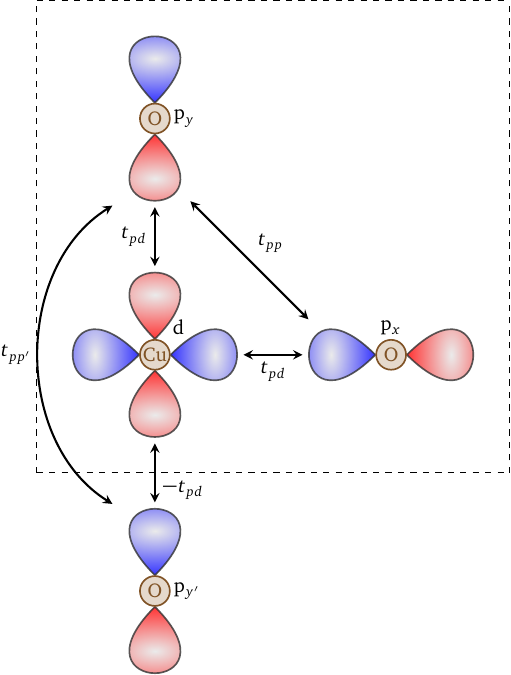}
    \caption{Schematic representation of the Emery model in position space. The orbitals are real and the signs of their lobes is indicated by the color. This is the gauge chosen for our basis states. The unit cell is the square delimited by dashed lines. $t_{pd}$ is the hopping amplitude between a $d$ and a $p$ orbital, $t_{pp}$ is the hopping amplitude between a $p_x$ and a $p_y$ orbital, $t_{pp'}$ is a hopping amplitude between two $p_x$ or two $p_y$ orbitals across a copper, $\epsilon_d$ is the on-site energy of $d$ orbitals, and $\epsilon_p$ is the on-site energy of $p$ orbitals. The on-site interaction $U_d$ is on the copper only.}
    \label{fig:emery_model}
\end{figure}

\section{Application to the Emery model}
\label{sec:Emery}
In section~\ref{sec:hamiltonian} we discuss the non-interacting part of the Hamiltonian. The choice of interaction and its consequence on the self-energy is in section~\ref{sec:self_energy_emery} while the effective one-band model that can be obtained by tracing out the oxygen orbitals is in section~\ref{sec:Effective_one_band}.
\subsection{Hamiltonian}
\label{sec:hamiltonian}
In the Emery model, we consider a bidimensional square lattice where each unit cell (noted with the index $i$ below) contains three distinct orbitals: one correlated $d$ orbital, and two noninteracting orbitals $p_x$ and $p_y$. The associated annihilation (creation) operators will be noted $d^{(\dagger)}_{i\sigma}$, $p^{(\dagger)}_{x,i\sigma}$ and $p^{(\dagger)}_{y,i\sigma}$ in real-space, with $\sigma$ a spin index. In reciprocal space, the unit cell index $i$ is replaced with a momentum index $\mathbf{k}$ so that the Hamiltonian becomes

\begin{equation}
    H = \sum_{\mathbf{k},\sigma}C^{\dagger}_{\mathbf{k}\sigma}\mathbf{h}^{(0)}(\mathbf{k})C_{\mathbf{k}\sigma} + U_d\sum_{i}n_{di\uparrow}n_{di\downarrow},
\end{equation}
where $C^{(\dagger)}_{\mathbf{k}\sigma} = (d^{(\dagger)}_{\mathbf{k}\sigma}, p^{(\dagger)}_{x,\mathbf{k}\sigma}, p^{(\dagger)}_{y,\mathbf{k}\sigma})$ is a vector of annihilation (creation) operators, $n_{\alpha i\sigma}$ is the number operator for orbital $\alpha$ on unit cell $i$, and $\mathbf{h}^{(0)}(\mathbf{k})$ is a $3\times 3$ matrix. This noninteracting part of the Hamiltonian is, in units where the lattice spacing is $a=1$ \cite{Fratino_2016,Kowalski_2021}\footnote{The gauge transformation $p_x \rightarrow -ie^{ik_x/2}p_x$, $p_y \rightarrow ie^{ik_y/2}p_y$ on this complex Hermitian Hamiltonian leads to the real Hamiltonian in Ref.~\citenum{Weber_2010}}
\begin{widetext}
    \begin{equation}
    \mathbf{h}^{(0)}(\mathbf{k}) = \begin{pmatrix} 
        \epsilon_d & t_{pd}(1-e^{-ik_x}) & t_{pd}(1-e^{-ik_y})\\
        t_{pd}(1-e^{ik_x}) & \epsilon_p + 2t_{pp'}\cos(k_x) & t_{pp}(1-e^{ik_x})(1-e^{-ik_y})\\
        t_{pd}(1-e^{ik_y}) & t_{pp}(1-e^{-ik_x})(1-e^{ik_y}) & \epsilon_p + 2t_{pp'}\cos(k_y)
    \end{pmatrix}, 
\end{equation}
\end{widetext}
where $t_{pd}$ is the hopping amplitude between a $d$ and a $p$ orbital, $t_{pp}$ is the hopping amplitude between a $p_x$ and a $p_y$ orbital, $t_{pp'}$ is a hopping amplitude between two $p_x$ or two $p_y$ orbitals across a copper, $\epsilon_d$ is the on-site energy of $d$ orbitals, and $\epsilon_p$ is the on-site energy of $p$ orbitals. In \fref{fig:emery_model}, we show a schematic representation of the Emery model and of its various parameters including the gauge choice.

\subsection{Interaction and self-energy for the Emery model}
\label{sec:self_energy_emery}
In the Emery model, the on-site interaction takes the form 
%%%%%%%%%%%%%%%%
\begin{equation}
    U_{\alpha\beta}=U_{d}\delta_{d\alpha}\delta_{d\beta}.
\end{equation}
%%%%%%%%%%%%%%%%
Using this in equations \ref{eq:selfEnergy_gen} and \ref{eq:selfEnergy_gen_tau2} for the self-energy, we obtain
%%%%%%%%%%%%%%%%
\begin{widetext}
\begin{align}
    \Sigma_{\alpha\bar{\gamma}\sigma}(1,\bar{3})_\phi \mathcal{G}_{\bar{\gamma}\beta\sigma}(\bar{3},2)_\phi &= -U_{d}\delta_{d\alpha}\delta_{d\bar{\gamma}}\left \langle T_\tau c^\dagger_{\bar{\gamma}-\sigma}(1^+)c_{\bar{\gamma}-\sigma}(1)c_{\alpha\sigma}(1)c^\dagger_{\beta\sigma}(2)\right \rangle_\phi,\label{eq:selfEnergyG_Emery_tau1}\\
    \Rightarrow \Sigma_{\alpha\beta\sigma}(1,2)_\phi &= -U_{d}\delta_{d\alpha}\left \langle T_\tau c^\dagger_{d-\sigma}(1^+)c_{d-\sigma}(1)c_{\alpha\sigma}(1)c^\dagger_{\bar{\gamma}\sigma}(\bar{4})\right \rangle_\phi \mathcal{G}_{\bar{\gamma}\beta\sigma}(\bar{4},2)_\phi^{-1}.
    \label{eq:selfEnergy_Emery_tau1}
\end{align}
%%%%%%%%%%%%%%%%
%%%%%%%%%%%%%%%%
\begin{align}
    \mathcal{G}_{\alpha \bar{\gamma}\sigma}(1,\bar{3})_\phi \Sigma_{\bar{\gamma}\beta\sigma}(\bar{3},2)_\phi &= -U_{d}\delta_{d\beta}\delta_{d\bar{\gamma}}\left \langle T_\tau c^\dagger_{\bar{\gamma}-\sigma}(2^+)c_{\bar{\gamma}-\sigma}(2)c_{\alpha\sigma}(1)c^\dagger_{\beta\sigma}(2)\right \rangle_\phi, \label{eq:selfEnergyG_Emery_tau2}\\
    \Rightarrow \Sigma_{\alpha\beta\sigma}(1,2)_\phi &= -U_{d} \delta_{d\beta} \mathcal{G}_{\alpha\bar{\gamma}\sigma}(1,\bar{4})_\phi^{-1} \left \langle T_\tau c^\dagger_{d-\sigma}(2^+)c_{d-\sigma}(2)c_{\bar{\gamma}\sigma}(\bar{4})c^\dagger_{\beta\sigma}(2)\right \rangle_\phi.
    \label{eq:selfEnergy_Emery_tau2}
\end{align}
\end{widetext}
%%%%%%%%%%%%%%%%
Since equations \ref{eq:selfEnergy_Emery_tau1} and \ref{eq:selfEnergy_Emery_tau2} should be satisfied at the same time, our first result for the Emery model is that the only non-zero element of the self-energy matrix is the one corresponding to the $d$ orbitals 
%%%%%%%%%%%%%%%%
\begin{equation}
    \Sigma_{\alpha\beta\sigma}(1,2)\propto \delta_{\alpha\beta}\delta_{\alpha d}.
\end{equation}
%%%%%%%%%%%%%%%%
The vertices hence become
%%%%%%%%%%%%%%%%
\begin{align}
    \Gamma^{\pm}_{\bar{\xi}\bar{\eta}\bar{\mu}\bar{\nu}}(\bar{6},\bar{5},\bar{7},\bar{8})_\phi &= \frac{\delta \Sigma_{\bar{\eta}\bar{\xi}\downarrow}(\bar{5},\bar{6})_\phi}{\delta \mathcal{G}_{\bar{\mu}\bar{\nu}\uparrow}(\bar{7},\bar{8})_\phi} \pm \frac{\delta \Sigma_{\bar{\eta}\bar{\xi}\uparrow}(\bar{5},\bar{6})_\phi}{\delta \mathcal{G}_{\bar{\mu}\bar{\nu}\uparrow}(\bar{7},\bar{8})_\phi},\nonumber \\
    &= \delta_{\bar{\eta}d}\delta_{\bar{\xi}d} \left ( \frac{\delta \Sigma_{dd\downarrow}(\bar{5},\bar{6})_\phi}{\delta \mathcal{G}_{\bar{\mu}\bar{\nu}\uparrow}(\bar{7},\bar{8})_\phi} \pm \frac{\delta \Sigma_{dd\uparrow}(\bar{5},\bar{6})_\phi}{\delta \mathcal{G}_{\bar{\mu}\bar{\nu}\uparrow}(\bar{7},\bar{8})_\phi} \right ).
    \label{eq:TPSC_vertices_1}
\end{align}
%%%%%%%%%%%%%%%%

\subsection{Effective one-band problem}
\label{sec:Effective_one_band}
Since only the $d$ orbitals are correlated in the Emery model, it is possible to cast it into an effective one-band model. Detailed calculations can be found in Appendix \ref{app:effective_one_band}. We first note that, to simplify the notation, we use the following matrix notation in orbital space, dropping spin indices
%%%%%%%%%%%%%%%%
\begin{align}
    \mathbf{A} &= \begin{pmatrix}
        A_{dd} & A_{dp_x} & A_{dp_y}\\
        A_{p_xd} & A_{p_xp_x} & A_{p_xp_y}\\
        A_{p_yd} & A_{p_yp_x} & A_{p_yp_y}
    \end{pmatrix},\\
    &\equiv \begin{pmatrix}
        A_{d} & \mathbf{A}_{dp} \\
        \mathbf{A}_{pd} & \mathbf{A}_{p} 
    \end{pmatrix}.
\end{align}
%%%%%%%%%%%%%%%%

In matrix form, where the matrix elements are orbital components, the Dyson equation is
%%%%%%%%%%%%%%%%
\begin{equation}
    \mathbf{G}(k)^{-1} =  \mathbf{G}^{(0)}(k)^{-1} - \mathbf{\Sigma}(k),
\end{equation}
%%%%%%%%%%%%%%%%
where we use the notation $(k)\equiv(\mathbf{k},ik_n)$, with $ik_n$ a fermionic Matsubara frequency. The noninteracting Green's function matrix is
%%%%%%%%%%%%%%%%
\begin{equation}
    \mathbf{G}^{(0)}(k)^{-1} = (ik_n+\mu)\mathbf{1}-\mathbf{h}^{(0)}(\mathbf{k}).
\end{equation}
%%%%%%%%%%%%%%%%
As seen previously, the self-energy in the Emery model is nonzero only for the $d$ orbital elements: $[\mathbf{\Sigma}(k)]_{\alpha\beta}=\Sigma_d(k)\delta_{\alpha d}\delta_{\beta d}$.

We obtain the following expression for the Green's function of the $d$ orbitals
%%%%%%%%%%%%%%%%
\begin{equation}
    \mathcal{G}_d(k) = \left [ ik_n+\mu-h^{(0)}_d(\mathbf{k})-\Sigma_d(k) - \Delta_{dp}(\mathbf{k},ik_n) \right ]^{-1},
\end{equation}
%%%%%%%%%%%%%%%%
where $\Delta_{dp}$ is a hybridization function between the $p$ and $d$ orbitals, defined as
%%%%%%%%%%%%%%%%
\begin{equation}
    \Delta_{dp}(k) \equiv \mathbf{h}^{(0)}_{dp}(\mathbf{k})\left ((ik_n+\mu)\mathbf{1}-\mathbf{h}_p^{(0)}(\mathbf{k})\right )^{-1}\mathbf{h}^{(0)}_{pd}(\mathbf{k}). 
\end{equation}
%%%%%%%%%%%%%%%%

Hence, the Emery model can be cast as an effective one-band problem. In principle, it can be solved by usual many-body methods for the Hubbard model, provided that the hybridization function is included in the calculation. 

% TPSC
\section{TPSC+DMFT approach for the Emery model}
\label{sec:TPSC+DMFT}
In TPSC, there are two steps (or levels of approximation). The calculation of the spin and charge susceptibilities and corresponding vertices are obtained first, as discussed in section~\ref{sec:first_level_ansatz}. The frequency-dependent self-energy is obtained at the second level of approximation in section~\ref{sec:second_level_self}. The calculations are done at constant total density at all stages, as explained in section~\ref{sec:mu_n}. The method and equations to be solved are summarized in section~\ref{sec:Methodology_TPSC+DMFT}.
\subsection{First level of approximation for the self-energy and TPSC ansatz}
\label{sec:first_level_ansatz}
Following the original formulation of the TPSC approach, we start by recalling the form of the self-energy within the Emery model. We first consider the self-energy obtained by the equation of motion with $\tau_1$, \eref{eq:selfEnergyG_Emery_tau1}. We obtain the first level of approximation by postulating a Hatree-Fock like decoupling in \eref{eq:selfEnergyG_Emery_tau1}
%%%%%%%%%%%%%%%%
\begin{align}
    \Sigma^{(1)}_{\alpha\bar{\gamma}\sigma}&(1,\bar{3})_\phi \mathcal{G}^{(1)}_{\bar{\gamma}\beta\sigma}(\bar{3},2\neq1)_\phi = \nonumber \\
    &A^{\alpha\beta}_\phi \delta_{d\alpha}\mathcal{G}^{(1)}_{\alpha\alpha-\sigma}(1,1^+)_\phi\mathcal{G}^{(1)}_{\alpha\beta\sigma}(1,2)_\phi,
    \label{eq:firstlevel}
\end{align}
%%%%%%%%%%%%%%%%
where the superscript $(1)$ denotes the first level of approximation. We obtain an expression for the self-energy by multiplying each side by an inverse Green's function
%%%%%%%%%%%%%%%%
\begin{equation}
    \Sigma_{\alpha\beta\sigma}^{(1)}(1,2)_\phi = A_\phi^{\alpha\beta}\delta_{d\alpha}\mathcal{G}^{(1)}_{\alpha\alpha-\sigma}(1,1^+)_\phi\delta_{\alpha \beta}\delta(1-2).
\end{equation}
%%%%%%%%%%%%%%%%
We can now evaluate the vertex $\Gamma^{-}_{\beta\alpha\delta\gamma}$ with this first approximation of the self-energy
%%%%%%%%%%%%%%%%
\begin{align}
    \Gamma^{-}_{\beta\alpha\delta\gamma}&(2,1;,3,4)_\phi = \frac{\delta \Sigma_{\alpha\beta\sigma}^{(1)}(1,2)_\phi}{\delta \mathcal{G}_{\delta\gamma-\sigma}^{(1)}(3,4)_\phi } -  \frac{\delta \Sigma_{\alpha\beta\sigma}^{(1)}(1,2)_\phi}{\delta \mathcal{G}_{\delta\gamma\sigma}^{(1)}(3,4)_\phi }, \nonumber \\
    &=  A_\phi^{\alpha\beta} \delta_{\alpha\delta}\delta_{\beta\gamma}\delta(3-1)\delta(4-1)\delta_{\alpha \beta}\delta(1-2)\delta_{d\alpha}.
\end{align}
%%%%%%%%%%%%%%%%

In the original TPSC approach, the functional derivatives of $A_\phi^{\alpha\beta}$ with respect to $\mathcal{G}_{\delta\gamma-\sigma}^{(1)}(3,4)_\phi$ and to $\mathcal{G}_{\delta\gamma\sigma}^{(1)}(3,4)_\phi$ cancel each other out. We find that, here also, this vertex is local and that its only non-zero component is for the $d$-orbital. We obtain an expression for the spin vertex by setting the source field to $0$
%%%%%%%%%%%%%%%%
\begin{equation}
    \Gamma^{-}_{\beta\alpha\delta\gamma}(2,1;,3,4)_\phi = U_{d}^{\mathrm{sp}}\delta_{d\alpha} \delta_{d\beta}\delta_{d\gamma}\delta_{d \delta} \delta(3-1)\delta(4-1)\delta(2-1).
\end{equation}
%%%%%%%%%%%%%%%%
The exact same result can be obtained using the self-energy defined from the derivative with respect to $\tau_2$ (\eref{eq:selfEnergy_gen_tau2}) instead of $\tau_1$ (\eref{eq:selfEnergy_gen}).

Following the original TPSC approach for the one-band Hubbard model, we obtain the value of the spin vertex $U_d^{\mathrm{sp}}$ by enforcing the local spin susceptibility sum rule. To do so, we return to the susceptibilities $\chi^{\pm}$ introduced in \sref{sec:chis_gammas_nn} and evaluate them with the first level of approximation for the Green's function, self-energy and vertices. We assume that the charge vertex is also local. In that case, \eref{eq:gen_chipm} for the generalized susceptibilities becomes
\begin{align}
    \chi^{\pm}_{\alpha\beta\gamma\delta}&(1,2;3,4) = \chi^{(1)}_{\alpha\beta\gamma\delta}(1,2;3,4)\nonumber \\
    &\mp \frac{1}{2} \chi^{(1)}_{\alpha\beta\bar{\xi}\bar{\eta}}(1,2;\bar{5},\bar{5^+})\Gamma^{\mathrm{ch,sp}}_{\bar{\xi}\bar{\eta}\bar{\mu}\bar{\nu}}\chi^{\pm}_{\bar{\mu}\bar{\nu}\gamma\delta}(\bar{5},\bar{5^+};3,4).
    \label{eq:gen_chipm_local}
\end{align}

In that equation and in the following, we use the notation $\chi^{(1)}$ to specify that the Green's functions used in the calculation of this bubble susceptibility  are evaluated at the first level of approximation: 
\begin{equation}
    \chi^{(1)}_{\alpha\beta\gamma\delta}(1,2;3,4) \equiv -2\mathcal{G}^{(1)}_{\alpha\delta\sigma}(1,4)\mathcal{G}^{(1)}_{\gamma\beta\sigma}(3,2).
    \label{eq:chi1def}
\end{equation}
 
The spin and charge susceptibilities are obtained by taking the limits $2\rightarrow 1^+$ and $4\rightarrow 3^+$ in the general RPA-like equation \eref{eq:gen_chipm_local}. Moreover, we simplify the notation with the following definitions:
\begin{align}
    \chi^{\pm}_{\alpha\beta\delta\gamma}(1,1^+;2,2^+) &\equiv \chi^{\mathrm{ch,sp}}_{\alpha\beta\delta\gamma}(1,2),\\
    \chi^{(1)}_{\alpha\beta\delta\gamma}(1,1^+;2,2^+) &\equiv \chi^{(1)}_{\alpha\beta\delta\gamma}(1,2).
\end{align}

Hence, the spin and charge susceptibilities are
\begin{equation}
    \chi^{\mathrm{ch,sp}}_{\alpha\beta\delta\gamma}(1,2) = \chi^{(1)}_{\alpha\beta\delta\gamma}(1,2) \mp \frac{1}{2} \chi^{(1)}_{\alpha\beta\bar{\xi}\bar{\eta}}(1,\bar{3})\Gamma^{\mathrm{ch,sp}}_{\bar{\xi}\bar{\eta}\bar{\mu}\bar{\nu}}\chi^{\mathrm{ch,sp}}_{\bar{\mu}\bar{\nu}\delta\gamma}(\bar{3},2).
\end{equation}

Performing a Fourier transform to momentum and Matsubara frequency space, using the notation $q\equiv(\mathbf{q},iq_n)$, leads to
\begin{equation}
    \chi^{\mathrm{ch,sp}}_{\alpha\beta\delta\gamma}(q) = \chi^{(1)}_{\alpha\beta\delta\gamma}(q) \mp \frac{1}{2} \chi^{(1)}_{\alpha\beta\bar{\xi}\bar{\eta}}(q)\Gamma^{\mathrm{ch,sp}}_{\bar{\eta}\bar{\xi}\bar{\nu}\bar{\mu}}\chi^{\mathrm{ch,sp}}_{\bar{\mu}\bar{\nu}\delta\gamma}(q).
    \label{eq:chichsp_q}
\end{equation}

Since the self-energy is only for the $d$ orbital, we assume, as in the spin case, that the charge vertex acts only on $d$ orbitals so that the susceptibilities obey
\begin{equation}
    \chi^{\mathrm{ch,sp}}_{\alpha\beta\delta\gamma}(q) = \chi^{(1)}_{\alpha\beta\delta\gamma}(q) \mp \frac{1}{2} \chi^{(1)}_{\alpha\beta dd}(q)U^{\mathrm{ch,sp}}_{d}\chi^{\mathrm{ch,sp}}_{dd\delta\gamma}(q).
\end{equation}

The charge and spin susceptibilities of the $d$ orbitals are hence fully decoupled from the other matrix elements 
\begin{equation}
    \chi^{\mathrm{ch,sp}}_{d}(q) = \chi^{(1)}_{d}(q) \mp \frac{1}{2} \chi^{(1)}_{d}(q)U^{\mathrm{ch,sp}}_{d}\chi^{\mathrm{ch,sp}}_{d}(q),
\end{equation}
\begin{equation}
    \Rightarrow \chi^{\mathrm{ch,sp}}_{d}(q) = \frac{\chi^{(1)}_{d}(q)}{1\pm  \frac{1}{2}U^{\mathrm{ch,sp}}_{d} \chi^{(1)}_{d}(q)}.
\end{equation}

We obtain the local spin and charge sum rules for the $d$-orbital susceptibilities from this last equation. Enforcing the Pauli principle, these sum rules are
\begin{equation}
    \frac{T}{N}\sum_{q}\frac{\chi^{(1)}_d(q)}{1-\frac{1}{2}U^{\mathrm{sp}}_d\chi_d^{(1)}(q)} = \langle n_d\rangle - 2\langle n_{d\uparrow}n_{d\downarrow}\rangle,
    \label{eq:sumrule_chisp}
\end{equation}

\begin{equation}
    \frac{T}{N}\sum_{q}\frac{\chi^{(1)}_d(q)}{1+\frac{1}{2}U^{\mathrm{ch}}_d\chi_d^{(1)}(q)} = \langle n_d\rangle + 2\langle n_{d\uparrow}n_{d\downarrow}\rangle - \langle n_d\rangle^2.
    \label{eq:sumrule_chich}
\end{equation}
We can use these sum rules to compute the values of the spin and charge vertices $U_d^{\mathrm{sp}}$ and $U_{d}^{\mathrm{ch}}$. 

The first option to do so is to use the original TPSC ansatz to enforce that the exact Migdal-Galitskii result
\begin{equation}
    \Sigma_{\alpha\bar{\gamma}\sigma}(1,\bar{3}) \mathcal{G}_{\bar{\gamma}\alpha\sigma}(\bar{3},1^+) = U_{d}\delta_{d\alpha}\left \langle T_\tau n_{\alpha\alpha-\sigma}(1)n_{\alpha\alpha\sigma}(1)\right \rangle
\end{equation}
is satisfied by the Hartree-like decoupling of the first-level self-energy \eref{eq:firstlevel}  evaluated at equal times ($\tau_2 = \tau_1^+$). 
For this equation to be satisfied, the spin vertex must take the form
\begin{equation}
    U^{\mathrm{sp}}_{d} = U_{d}\frac{\left \langle n_{d\uparrow}n_{d\downarrow}\right \rangle}{\langle n_{d\uparrow}\rangle \langle n_{d\downarrow}\rangle}.
    \label{eq:ansatz_usp}
\end{equation}
A solution for the spin vertex and the double occupancy is then obtained in a self-consistent way by solving simultaneously the {\it ansatz}~\eref{eq:ansatz_usp} and the sum rule for the spin susceptibility. We note that equation \ref{eq:ansatz_usp} does not satisfy particle-hole symmetry. In the one-band case, even when the hamiltonian itself does not satisfy particle-hole symmetry, the most reliable way to obtain the spin vertex when the density is larger than $1$ (electron doping) is to use a symmetrical ansatz of the form
\begin{equation}
    U^{\mathrm{sp}} =\left\{\begin{matrix} U\frac{\left \langle n_{\uparrow}n_{\downarrow}\right \rangle}{\langle n_{\uparrow}\rangle \langle n_{\downarrow}\rangle} & & n <1,\\
    &&\\
    U\frac{\left \langle (1-n_{\uparrow})(1-n_{\downarrow})\right \rangle}{\langle (1-n_{\uparrow})\rangle \langle (1-n_{\downarrow})\rangle} & & n >1.
    \end{matrix}\right.
    \label{eq:ansatz_sym}
\end{equation}
We use the same symmetrical ansatz \eref{eq:ansatz_sym} for the Emery model, with the $d$-orbital densities $n_{d\sigma}$.  

The second option is to use the TPSC+DMFT approach introduced in Ref.~\cite{Martin_2023} for the one-band Hubbard model. In this approach, the values of the spin and charge vertices are still obtained by satisfying the sum rules. However, instead of using the ansatz \eref{eq:ansatz_usp}, the double occupancy used for the computation of the sum rules is obtained by a DMFT calculation. This method can also be applied to the Emery model since, as mentioned earlier, this model can be mapped to an effective one-band model that can be solved by single-site DMFT calculations. The TPSC+DMFT formalism was also applied in a different formulation of the multiorbital TPSC approach in Ref. \cite{Zantout_2022}. 

\subsection{Second level of approximation for the self-energy}
\label{sec:second_level_self}
In TPSC, we obtain a better approximation for the self-energy at the second level of approximation. 
We concluded in section \ref{sec:first_level_ansatz} that the TPSC irreducible spin vertex matrix is nonzero only for the $d$ orbitals, and in section \ref{sec:self_energy_emery} that this is also true for the self-energy matrix. Hence, in the TPSC approach for the Emery model, the self-energy at the second level of approximation follows the same crossing-symmetric equation as in the one-band Hubbard model \cite{Moukouri_2000,Allen_2003}, namely
\begin{align}
    \Sigma^{(2)}_{d}(k) &= U_d\frac{\langle n_d \rangle}{2} + \frac{U_d}{8}\frac{T}{N}\sum_{q}\left[ 3U^{\mathrm{sp}}_{d}\chi^{\mathrm{sp}}_{d}(q)\right. \nonumber \\
    & \left .+ U^{\mathrm{ch}}_{d}\chi^{\mathrm{ch}}_{d}(q) \right]\mathcal{G}_d^{(1)}(k+q).
    \label{eq:self_energy_2}
\end{align}

\subsection{Chemical potentials and densities}
\label{sec:mu_n}
Having developed an effective one-band model, it would be tempting to follow the TPSC formalism formulated for the one-band Hubbard model. The steps necessary to solve the first level of approximation within the original TPSC are:
%%%%%%%%%%%%%%%%
\begin{enumerate}
    \item Calculate the Green's function at the first level of approximation from an input total density,
    \item Calculate the noninteracting correlation function $\chi^{(1)}$,
    \item Calculate the spin vertex $U^{\mathrm{sp}}_d$ from the local sum rule for the spin susceptibility and the TPSC ansatz or the DMFT double occupancy,
    \item Calculate the charge vertex $U^{\mathrm{ch}}_d$ from the local sum rule for the spin susceptibility and the TPSC ansatz or the DMFT double occupancy.
\end{enumerate}
%%%%%%%%%%%%%%%%
The first challenge we face in trying to follow these steps concerns the local spin and charge sum rules defined in \eref{eq:sumrule_chisp} and \eref{eq:sumrule_chich}.

In the one-band case, we take the density $n$ as an input and calculate the chemical potential so that the Green's function at the first level of approximation $\mathcal{G}^{(1)}$ returns the right density. The self-energy at the first level of approximation being a constant, it can be absorbed in the definition of the chemical potential of $\mathcal{G}^{(1)}$. Previous formulations of the TPSC approach for multiorbital Hamiltonians have kept the same methodology, computing the sum rules with a noninteracting density for orbital $\alpha$ $\langle n_\alpha\rangle_0$ that is obtained from an input total density $n$. Mathematically, this means that the chemical potential is first calculated from the total density from the equation
\begin{equation}
    n = \frac{1}{N}\sum_{\mathbf{k},\alpha} \mathcal{G}_{\alpha\alpha}^{(0)}(\mathbf{k},\tau=0^-),
\end{equation}
where the noninteracting Green's function follows
\begin{equation}
    \mathcal{G}_{\alpha\alpha}^{(0)}(\mathbf{k},ik_n)^{-1}= ik_n+\mu^{(0)}-h^{(0)}_{\alpha\alpha}(\mathbf{k}).
\end{equation}
Then, the noninteracting density $\langle n_\alpha\rangle_0$ is obtained from
\begin{equation}
    \langle n_\alpha\rangle_0 =\frac{1}{N}\sum_{\mathbf{k}}  \mathcal{G}_{\alpha\alpha}^{(0)}(\mathbf{k},\tau=0^-).
\end{equation}

Hence, in these formulations, the \textit{total} density  is kept constant and is used to compute the chemical potential of a noninteracting Green's function $\mathcal{G}^{(0)}$, from which the orbital-resolved densities are deduced. This approach should work in the specific case of degenerate orbitals. However, this approximation is not valid in general due to two (related) reasons:

\textbf{1 - Noninteracting vs first level of approximation} In the TPSC approach for the one-band Hubbard model, the self-energy at the first level of approximation is a constant that is usually absorbed in the definition of the chemical potential. In the previous section, we saw that the self-energy $\Sigma^{(1)}_{\alpha\beta}$ in a multiorbital case is a constant in the sense that it does not depend on the momentum or the frequency. However, it does depend on the orbital indices. This should be true in general for multiorbital Hamiltonians. Since this self-energy is orbital-dependent, the usual TPSC-way of computing the susceptibilities with the noninteracting Green's function containing an adjusted chemical potential should not be valid anymore (except for degenerate orbitals). More specifically, computing a single chemical potential from a total density at the ``zeroth'' level amounts to taking the same constant  self-energy for all orbitals: $\Sigma^{(1)}_{\alpha\alpha} = C~\forall~\alpha$. Such an approximation is not valid for nondegenerate models such as the Emery model.

\textbf{2 - Noninteracting vs interacting orbital densities} The densities that appear in the sum rules for the spin and charge susceptibilities should be calculated in the presence of interaction. In the case of degenerate orbitals, we can expect that these values are the same: $\langle n_\alpha\rangle_0 = \langle n_\alpha\rangle$. However, this is not the case for the Emery model \cite{Fratino_2016}.

In other words, in multiorbital models the Green's function that should be used at the first level of approximation should be computed from interacting densities $\langle n_\alpha\rangle$, by solving the set of equations
\begin{equation}
    \mathcal{G}_{\alpha\alpha}^{(1)}(\mathbf{k},ik_n)^{-1}= ik_n+\mu^{(1)}-h^{(0)}_{\alpha\alpha}(\mathbf{k})-\Sigma_{\alpha\alpha}^{(1)},
    \label{eq:set_equation_level1_1}
\end{equation}
\begin{equation}
    \langle n_\alpha\rangle =  \mathcal{G}_{\alpha\alpha}^{(1)}(\mathbf{k},\tau=0^-).
    \label{eq:set_equation_level1_2}
\end{equation}
Unfortunately, this system of equations has too many unknowns to be solved on its own: assuming we input a total density $n$, the orbital densities $\langle n_\alpha\rangle$, the chemical potential $\mu^{(1)}$ and the self-energies $\Sigma^{(1)}_{\alpha\alpha}$ are all unknown. This is resolved in the following section.

\subsection{Methodology for the TPSC+DMFT approach}
\label{sec:Methodology_TPSC+DMFT}
We circumvent the issue raised in the previous section by using the TPSC+DMFT approach developed in Ref. \cite{Martin_2023}. This approach, which we mentioned briefly in \sref{sec:first_level_ansatz}, follows the steps outlined below:
\begin{enumerate}
    \item For a given parameter set (hopping parameters, interaction strength, temperature), perform a DMFT calculation at a specific chemical potential $\mu_{\mathrm{DMFT}}$ corresponding to the desired total density,
    \item The DMFT calculation yields observables such as the total density $n$ and the double occupancy $\langle n_\uparrow n_\downarrow\rangle_{\mathrm{DMFT}}$,
    \item Perform a TPSC calculation with the same parameter set for the total density obtained with the DMFT calculation.
    \item Instead of using the TPSC ansatz, use the double occupancy from DMFT $\langle n_\uparrow n_\downarrow\rangle_{\mathrm{DMFT}}$ as an input to obtain the spin and charge vertices from the spin and charge sum rules.
\end{enumerate}

We adapt this method for the Emery model by using more inputs from DMFT than in the one-band case. More specifically, we need:
%%%%%%%%%%%%%%%%
\begin{itemize}
    \item The orbital densities of the $p$ and $d$ orbitals, $\langle n_p\rangle_{\mathrm{DMFT}}$ and $\langle n_d\rangle_{\mathrm{DMFT}}$,
    \item The double occupancy in the $d$ orbitals, $\langle n_{d\uparrow}n_{d\downarrow}\rangle$.
\end{itemize}
%%%%%%%%%%%%%%%%

\paragraph*{TPSC+DMFT for the Emery model --} In summary then, using as input densities those calculated for the interacting model with DMFT, we find the chemical potential and the self-energy at the first level of approximation by solving the set of equations (\ref{eq:set_equation_level1_1}-\ref{eq:set_equation_level1_2}). Then, we use again the input density $\langle n_d\rangle_{\mathrm{DMFT}}$ to calculate the spin and charge vertices from the sum rules Eqs.~\eqref{eq:sumrule_chisp} and~\eqref{eq:sumrule_chich} where the double occupancy $\langle n_{d\uparrow}n_{d\downarrow}\rangle_\mathrm{DMFT}$ on the right-hand side of these equations is also taken from DMFT. We can alternatively use the TPSC ansatz Eq.~\ref{eq:ansatz_sym} for double occupancy, as discussed in the following section. At this stage, the spin and charge susceptibilities have been calculated. The improved self-energy Eq.~\eqref{eq:self_energy_2} obtained at the second level of approximation follows from the vertices, susceptibilities and Green's functions computed at the first level.

% Résultats
\section{Results}
\label{sec:results}
In this section, we apply the method we outlined above to a specific realization of the Emery model. The goal of this section is not to provide specific, new physical insights into the model, but rather to compare the results obtained with the TPSC and TPSC+DMFT approaches to the Emery model. We consider the Emery model described in section \ref{sec:hamiltonian}. We set the hopping parameter $t_{pp}=1$ as the unit of energy. The other hopping parameters are chosen as $t_{pd}=2.15$, $t_{pp'}=0.2$, and the on-site energies as $\epsilon_d=-5.92$ and $\epsilon_p=0$. These parameters are inspired by the parameters for the electron-doped cuprate Nd$_{2-x}$Ce$_x$CuO$_4$ (NCCO) detailed in Ref. \cite{Weber_2010}. The value of the parameter $t_{pp'}$ is inspired by further work on the hole-doped cuprates \cite{Weber_2012} and again on NCCO \cite{Ogura_2015}. In the following, we take the Hubbard interaction strength on the $d$ orbitals as $U_d=10$, which is lower than the value suggested in Ref. \cite{Weber_2010} for NCCO ($U_d\simeq 14.82$). We choose this value of $U_d$ in order to stay away from the Mott transition. All our calculations are performed at the temperature $T=0.1$.

We start this section by discussing the difference between the interacting and noninteracting orbital densities in subsection \ref{sec:results_densities}. Then, we study the double occupancy and the irreducible spin vertex $U^{\mathrm{sp}}_d$ in subsection \ref{sec:results_docc_usp}, and the self-energy in subsection \ref{sec:results_self_energy}. In the subsection \ref{sec:local_self_energy} of our discussion for the self-energy, we include a comment on an additional step to the TPSC+DMFT method, which is to substitute the local part of the TPSC self-energy with the DMFT local self-energy. 

The single-site DMFT calculations are performed using the TRIQS library \cite{Parcollet_2015, Seth_2016}, with the CT-HYB quantum Monte Carlo impurity solver \cite{Werner_2006,Gull_2011}. The TPSC calculations are performed using the sparse-ir library \cite{Shinaoka_2017, Li_2020, Shinaoka_2022}. 

\subsection{Interacting vs noninteracting densities}
\begin{figure}
    \centering
    \includegraphics[width=0.8\columnwidth]{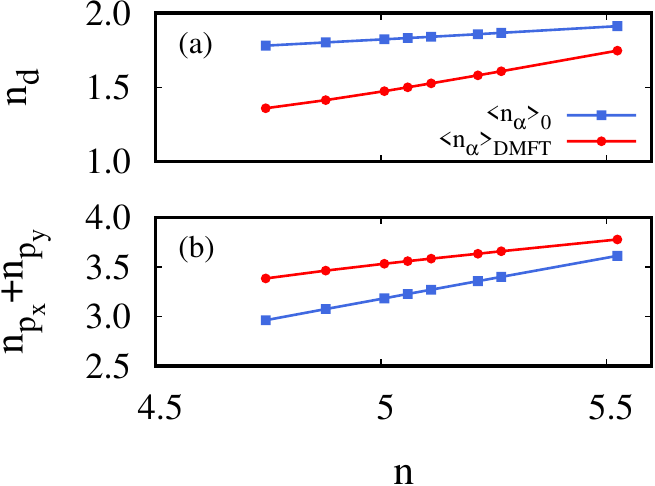}
    \caption{Orbital densities $\langle n_\alpha \rangle$ (a) of the $d$ orbitals ($\langle n_\alpha \rangle = \langle n_d \rangle $) and (b) of the $p$ orbitals ($\langle n_\alpha \rangle = \langle n_{p_x} + n_{p_y}\rangle$) as a function of the total density $n$. In blue, the densities are calculated from the noninteracting hamiltonian. In red, the densities are obtained from DMFT calculations with the interacting hamiltonian. The $d$-orbital density decreases when interactions are included, whereas the $p$-orbital density increases. The model parameters are $t_{pp}=1$, $t_{pd}=2.15$, $t_{pp'}=0.2$,  $\epsilon_d=-5.92$  $\epsilon_p=0$, $U_d=10$.}
    \label{fig:nd_np}
\end{figure}

We first calculate the densities with ($U_d=10$) and without ($U_d=0$) interactions with single-site DMFT. We show our results in Figure \ref{fig:nd_np}, where we illustrate (a) the $d$-orbital density and (b) the $p_x$- and $p_y$-orbital densities as a function of the total density $n$. We show that the $d$-orbital density decreases when the interaction $U_d$ is present, while the $p$-orbital densities increase. This is due to the asymmetry of the on-site energies $\epsilon_d\neq \epsilon_p$ as well as the on-site interaction $U_d$. Without interactions, the $p$-orbital densities are smaller than the $d$-orbital one since the on-site energy on the $d$ orbital is lower than on the $p$ orbitals. When the interactions are turned on, the on-site interaction $U$ leads to a decrease of the double occupancy on the $d$ orbitals, which in turn lowers the orbital occupancy on these orbitals. As a specific example, we consider the undoped case with total density $n=5$. In the noninteracting case, the orbital densities are $\langle n_d\rangle_0 \simeq 1.82$ and $\langle n_{p_x}+n_{p_y}\rangle_0 \simeq 3.18$. In contrast, in the interacting case, the orbital densities are  $\langle n_d\rangle_{\mathrm{DMFT}} \simeq 1.47$ and $\langle n_{p_x}+n_{p_y}\rangle_{\mathrm{DMFT}} \simeq 3.53$. It is self-evident that performing a TPSC calculation with the noninteracting density $\langle n_d\rangle_0 \simeq 1.82$ will yield results drastically different (incorrect) from those with the interacting density $\langle n_d\rangle_{\mathrm{DMFT}} \simeq 1.47$.
 
\label{sec:results_densities}

\subsection{Double occupancy and spin vertex}
\label{sec:results_docc_usp}

We now turn to the calculation of the double occupancy on the $d$ orbitals $D=\langle n_{d\uparrow} n_{d\downarrow} \rangle$. As mentioned in section \ref{sec:first_level_ansatz}, the double occupancy can be obtained either self-consistently with the TPSC ansatz or as an input from DMFT. Hence, we perform the TPSC calculation in three different ways: (1) from the noninteracting density $\langle n_d\rangle_0$ and the TPSC ansatz (Equation \ref{eq:ansatz_sym}), (2) from the interacting density obtained with DMFT $\langle n_d\rangle_{\mathrm{DMFT}}$ and the TPSC ansatz, and (3) with both the interacting density and the double occupancy obtained with DMFT, which will be noted $D_{\mathrm{DMFT}}$. As seen in the previous section, the $d$-orbital densities are larger than $1$ for all the total densities $n$ considered here, which means that all the calculations with the TPSC ansatz are performed with the particle-hole transformed ansatz.

\begin{figure}
    \centering
    \includegraphics[width=\columnwidth]{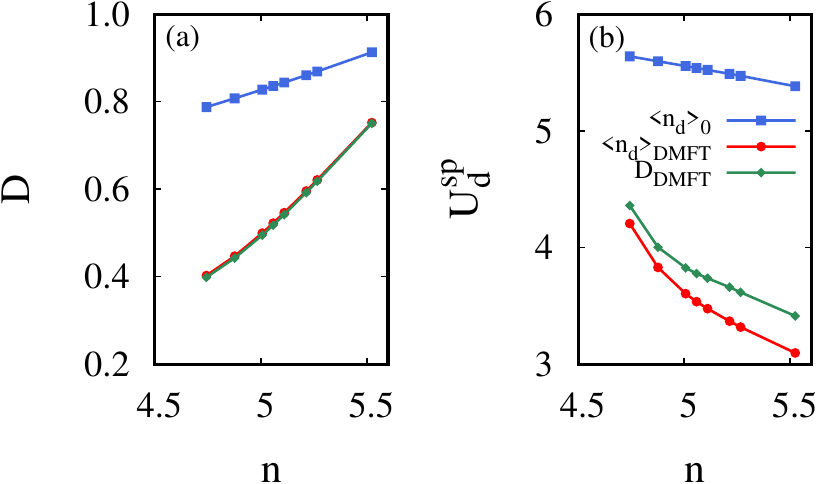}
    \caption{Effect of the input density on (a) the double occupancy and (b) the irreducible spin vertex $U^{\mathrm{sp}}_d$. In blue, the TPSC calculation is performed with the noninteracting orbital densities. In red, the TPSC calculation is performed with the interacting orbital densities from DMFT and the ansatz. In green, the TPSC calculation is performed with the interacting orbital densities and the double occupancy both from DMFT. The model parameters are $t_{pp}=1$, $t_{pd}=2.15$, $t_{pp'}=0.2$,  $\epsilon_d=-5.92$  $\epsilon_p=0$, $U_d=10$.}
    \label{fig:docc_usp}
\end{figure}

In Figure \ref{fig:docc_usp} (a), we show the double occupancy obtained from the three types of calculations listed above as a function of the total density $n$. The double occupancy obtained from the noninteracting density $\langle n_d\rangle_0$ and the TPSC ansatz is much higher than the ones obtained from the interacting density $\langle n_d\rangle_{\mathrm{DMFT}}$, in line with the fact that the noninteracting $d$-orbital density is larger than the one with interactions. We note that, in contrast, the double occupancy obtained with the TPSC ansatz and the interacting density is very close to the DMFT double occupancy, $D_{\mathrm{DMFT}}$, the two curves basically superposing on the plot. 

Next, we consider the irreducible spin vertex $U^{\mathrm{sp}}_d$ obtained from the three types of calculations. We show our results in Figure \ref{fig:docc_usp} (b). Despite the close agreement of the double occupancies obtained from the input DMFT densities with either the DMFT double occupancy or  the TPSC ansatz, we observe that the resulting values of $U^{\mathrm{sp}}_d$ for both calculations are distinct. The value of $U^{\mathrm{sp}}_d$ obtained with the interacting densities and the TPSC ansatz is systematically smaller than the one obtained with the interacting densities and the DMFT double occupancy, though both methods yield results that follow the same qualitative behavior. Again, the results obtained from the noninteracting densities are quantitatively quite different from the others, although the qualitative dependence on total density is similar.

Hence, in the model we consider here, the impacts of using the noninteracting densities rather than the interacting ones in the multiorbital TPSC calculations are higher than the impacts of using the double occupancy from DMFT or from the TPSC ansatz.

Note that our most reliable value of $U^\mathrm{sp}_d$ decreases rapidly with increasing filling, consistent with the idea that more electrons leads to better screening.
This decrease of $U^\mathrm{sp}_d$ leads to double-occupancy increasing with total density in Fig.~\ref{fig:docc_usp}(a) and to the interacting densities becoming closer to the noninteracting ones in Fig.~\ref{fig:nd_np}. 
This decrease of $U^\mathrm{sp}_d$ with filling is one of the factors that might explain why electron-doped cuprates appear less strongly correlated than their hole-doped counter-part, even though one expects the bare interaction $U$ to be comparable. 
Other factors, such as the absence of apical oxygen in electron-doped cuprates also contribute to this difference~\cite{Weber_2010}. In the one-band model, it was necessary to postulate a renormalization of the bare interaction $U$ to find agreement with ARPES experiments~\cite{Senechal:2004,Kyung_2004}.

\subsection{Self-Energy}
\label{sec:results_self_energy}
In this section, we calculate the self-energy at the second level of approximation of TPSC from equation \ref{eq:self_energy_2}. We start by discussing the impact of the input density. Next, we study the anisotropy of the self-energy with respect to the momentum on the Fermi surface. Finally, we consider an additional TPSC+DMFT step, where we substitute the local self-energy from DMFT in the TPSC self-energy.

\subsubsection{Impact of the input density}

As we did for the calculation of the double occupancy and the irreducible spin vertex, we start our discussion of the self-energy by comparing the results obtained from the three types of calculations described before. As a reminder, these are (1) with the noninteracting density $\langle n_d\rangle_0$ and the TPSC ansatz (Equation \ref{eq:ansatz_sym}), (2) with the interacting density obtained with DMFT $\langle n_d\rangle_{\mathrm{DMFT}}$ and the TPSC ansatz, and (3) with both the interacting density and the double occupancy obtained with DMFT, which is still noted $D_{\mathrm{DMFT}}$.

\begin{figure}
    \centering
    \includegraphics[width=\columnwidth]{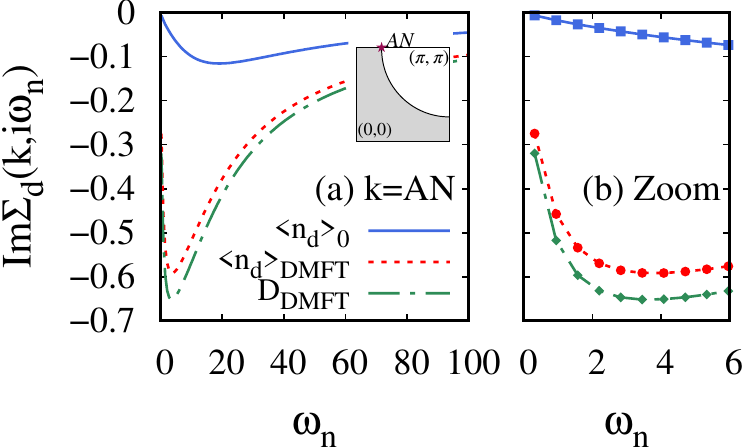}
    \caption{Effect of the input density on the self-energy, here shown at the antinodal wave-vector. Figure (a) shows the Matsubara frequency dependence of the self-energy, and figure (b) shows a zoom on the low Matsubara frequencies. These results are obtained at the total density $n=5.0069$. In blue, the TPSC calculation is performed with the noninteracting orbital densities. In red, the TPSC calculation is performed with the interacting orbital densities from DMFT and the ansatz. In green, the TPSC calculation is performed with the interacting orbital densities and the double occupancy both from DMFT. The model parameters are $t_{pp}=1$, $t_{pd}=2.15$, $t_{pp'}=0.2$,  $\epsilon_d=-5.92$  $\epsilon_p=0$, $U_d=10$.}
    \label{fig:selfEnergy_impact_density}
\end{figure}

To illustrate the impact of the type of method on the self-energy, we consider a specific case with a total density $n=5.0069$. We study the self-energy calculated at the antinodal point on the Fermi surface as a function of the Matsubara frequencies. The antinodal point is the wave vector where the Fermi surface crosses the Brillouin zone boundary edge: $\mathbf{k}_{\mathrm{AN}}=(k_{x,\mathrm{AN}},\pi)$ and is shown in the inset of Figure \ref{fig:selfEnergy_impact_density} (a) as a pink star.

As was the case in the calculation of the double occupancy and the spin vertex, the choice of the input density, whether it is the noninteracting or the interacting one, has the most visible effect on the self-energy. 

We can assess quantitatively the difference between the three approaches through the calculation of the quasiparticle weight defined as $Z_{\mathbf{k}} = \left (1-\left .\frac{\partial \mathrm{Re} \Sigma_d(\mathbf{k},\omega)}{\partial \omega}\right|_{\omega=0}\right)^{-1}$. We approximate the quasiparticle weight through a second order polynomial fit of the Matsubara self-energy over the first three Matsubara frequencies
\begin{equation}
    \mathrm{Im}\Sigma_d(\mathbf{k},i\omega_n) \simeq \alpha_0(T)+\alpha_1 \omega_n + \alpha_2 \omega_n^2,
\end{equation}
\begin{equation}
    Z_{\mathbf{k}} \simeq \frac{1}{1-\alpha_1}.
\end{equation}

This approximation is valid if the self-energy has a Fermi liquid form at low frequency. From this, we obtain the quasiparticle weights $Z_{\mathrm{AN}}\simeq 0.98$ for the calculation with the noninteracting density, $Z_{\mathrm{AN}}\simeq 0.68$ for the calculation with the interacting density and the TPSC ansatz, and $Z_{\mathrm{AN}}\simeq 0.66$ for the calculation with the interacting density and the DMFT double occupancy.  

\subsubsection{Nodal vs antinodal anisotropy}

\begin{figure}
    \centering
    \includegraphics[width=0.8\columnwidth]{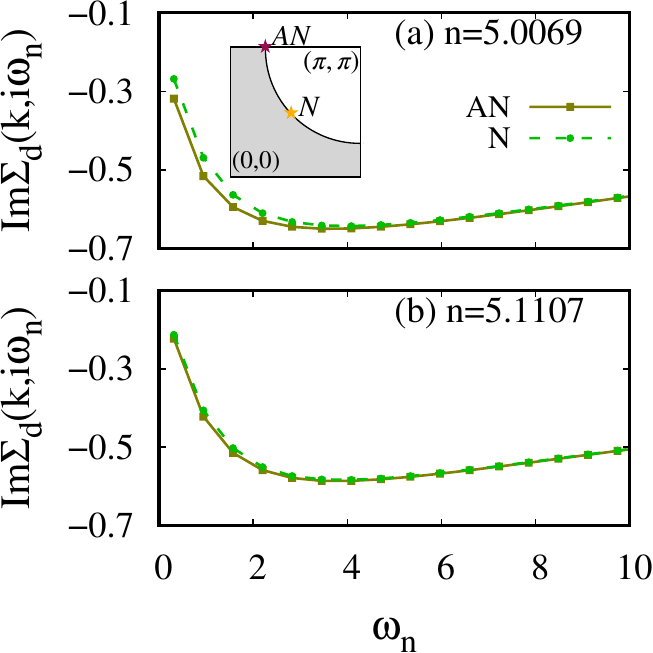}
    \caption{Imaginary part of the self-energy as a function of the Matsubara frequency. The results are shown for the antinodal wave-vector (squares, solid line), and the nodal wave-vector (circles, dashed line), for the densities (a) $n=5.0069$ and (b) $n=5.1107$. The results are obtained from TPSC calculations performed with the interacting orbital densities and the double occupancy both from DMFT. The model parameters are $t_{pp}=1$, $t_{pd}=2.15$, $t_{pp'}=0.2$,  $\epsilon_d=-5.92$  $\epsilon_p=0$, $U_d=10$.}
    \label{fig:selfEnergy_anisotropy}
\end{figure}

The self-energy obtained with TPSC is fully momentum-dependent, in contrast to the DMFT self-energy, which is purely local. The TPSC approach can hence be used to assess the anisotropy of the self-energy at different wave vectors on the Fermi surface. In figure \ref{fig:selfEnergy_anisotropy}, we illustrate this phenomenon by comparing the self-energy obtained at the antinodal and nodal points on the Fermi surface, indicated in the inset of the subfigure (a) by pink and yellow stars respectively. The results shown in this figure are obtained from the TPSC calculation using both the interacting density and the double occupancy from DMFT. In figure \ref{fig:selfEnergy_anisotropy} (a), the results are shown for the total density $n=5.0069$. In figure \ref{fig:selfEnergy_anisotropy} (b), the results are shown for the total density $n=5.1107$. At high frequency, for both densities, the antinodal and nodal self-energies coincide, as expected from previous analytical arguments \cite{Vilk_1997}. However, we note an anisotropy between the nodal and antinodal self-energies at low frequencies ($\omega_n<4$) for the density $n=5.0069$. This anisotropy also exists for the density $n=5.1107$, but to a lesser extent.

\begin{figure}
    \centering
    \includegraphics[width=\columnwidth]{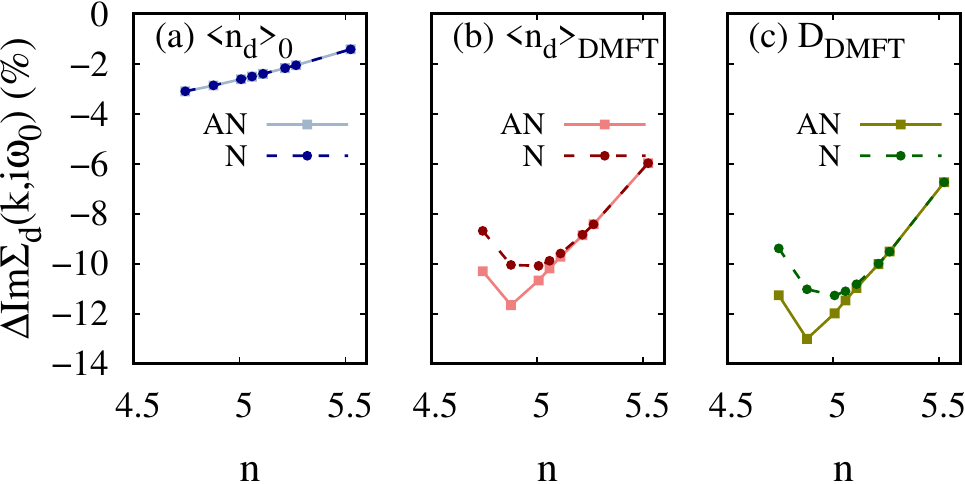}
    \caption{Relative deviation between the local part of the self-energy and the self-energy calculated at the first Matsubara frequency at the nodal (N, dashed lines with circles) and antinodal (AN, solid lines with squares) wave-vectors, as a function of the total density $n$. The deviation is calculated using equation \ref{eq:dev_sigma_loc_iw0}. The TPSC calculations are performed (a) with the noninteracting orbital densities, (b) with the interacting orbital densities from DMFT and (c) with both the interacting orbital densities and the double occupancy from DMFT. The model parameters are $t_{pp}=1$, $t_{pd}=2.15$, $t_{pp'}=0.2$,  $\epsilon_d=-5.92$  $\epsilon_p=0$, $U_d=10$.}
    \label{fig:selfEnergy_anisotropy_iw0}
\end{figure}

We obtain a quantitative view of the nodal and antinodal anisotropy by looking at the first Matsubara frequency, $\omega_0=\pi T$, where the anisotropy is the largest. In Fig.~\ref{fig:selfEnergy_anisotropy_iw0}, we show the relative deviation between the local self-energy and the self-energy at the antinodal and nodal points on the Fermi surface, defined by
\begin{equation}
    \Delta \mathrm{Im}\Sigma_d(\mathbf{k},i\omega_0) = \frac{\mathrm{Im}\Sigma_d(\mathbf{k},i\omega_0)-\mathrm{Im}\Sigma_{d, \mathrm{loc}}(i\omega_0)}{\mathrm{Im}\Sigma_{d, \mathrm{loc}}(i\omega_0)}.
    \label{eq:dev_sigma_loc_iw0}
\end{equation}
This relative deviation is shown in Fig.~\ref{fig:selfEnergy_anisotropy_iw0} as a function of the total density $n$ for TPSC calculations performed with (a) the noninteracting density and the TPSC ansatz, (b) the interacting density from DMFT and the TPSC ansatz and (c) the interacting density and the double occupancy both from DMFT.

In the calculations with the noninteracting density, the deviation is of the order of only $2$ to $4$\% (in absolute value) and is equal at the node and the antinode. This is an indication of an isotropic self-energy along the Fermi surface. 

In contrast, the deviation obtained for both types of calculations performed with the interacting density reach $\simeq 12$\% in absolute value around the density $n=5$, corresponding to the undoped case. The deviation from the local self-energy at the antinode and at the node is different, which is a clear indication of the anisotropy of the self-energy on the Fermi surface.

Using the TPSC approach for the Emery model hence leads to a visible anisotropy of the self-energy on the Fermi surface, which cannot be obtained by DMFT alone. However, in the model we consider, it is necessary to use the interacting density from DMFT to capture this anisotropy correctly.

\subsubsection{Additional TPSC+DMFT step: substitution of the local self-energy from DMFT}
\label{sec:local_self_energy}

In Ref. \cite{Martin_2023}, the TPSC+DMFT approach for the Hubbard model was introduced with two distinct steps. The first step, which we have considered so far, is to use the DMFT double occupancy to obtain the spin vertex $U^{\mathrm{sp}}_d$ rather than using the TPSC ansatz. We now consider the second step, which consists in substituting the local self-energy obtained from the TPSC calculation with the one obtained from DMFT
\begin{equation}
    \Sigma_d(\mathbf{k},i\omega_n) \rightarrow \Sigma^{\mathrm{TPSC}}_d(\mathbf{k},i\omega_n)-\Sigma^{\mathrm{TPSC}}_{d, \mathrm{loc}}(i\omega_n)+\Sigma^{\mathrm{DMFT}}_{d, \mathrm{loc}}(i\omega_n).
    \label{eq:sub_sigma_loc_dmft}
\end{equation}

\begin{figure}
    \centering
    \includegraphics[width=0.9\columnwidth]{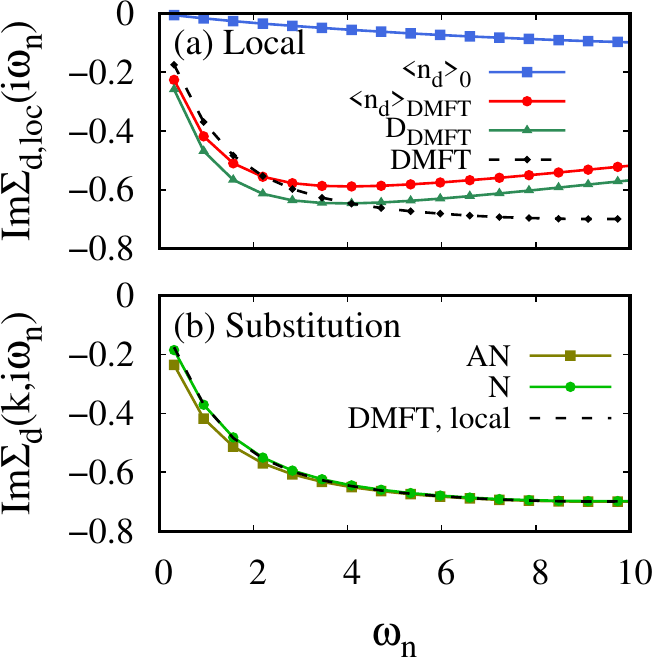}
    \caption{(a) Local part of the self-energy obtained from TPSC calculations (full lines) and from DMFT (black dashed line). In blue, the TPSC calculation is performed with the noninteracting orbital densities. In red, the TPSC calculation is performed with the interacting orbital densities from DMFT. In green, the TPSC calculation is performed with both the interacting orbital densities and the double occupancy from DMFT. (b) Imaginary part of the self-energy calculated at the nodal (N) and antinodal (AN) wave-vectors, where the local part from DMFT was substituted following equation \ref{eq:sub_sigma_loc_dmft}. For both subplots, the total density is $n=5.0069$. The model parameters are $t_{pp}=1$, $t_{pd}=2.15$, $t_{pp'}=0.2$,  $\epsilon_d=-5.92$  $\epsilon_p=0$, $U_d=10$.}
    \label{fig:selfEnergy_comp_DMFT}
\end{figure}

For the one-band Hubbard model, this approach was found to yield better results for the self-energy through benchmarks with exact diagrammatic Monte Carlo calculations \cite{Martin_2023}. Hence we explore this approach here for the Emery model, though we do not have exact benchmarks to compare with. We illustrate this process for the $n=5.0069$ case in Fig.~\ref{fig:selfEnergy_comp_DMFT}.

In Fig.~\ref{fig:selfEnergy_comp_DMFT}~(a), we compare the local self-energy obtained with DMFT (black dashed line) to the TPSC local self-energy. Once again, we compare  results obtained from TPSC calculations performed with the noninteracting density and the TPSC ansatz, the interacting density and the TPSC ansatz, and the interacting density and double occupancy both obtained from DMFT. This comparison shows clearly that using the interacting density in the TPSC calculation is necessary in order to obtain a local self-energy that is at least qualitatively correct when compared with DMFT. 

In Fig.~\ref{fig:selfEnergy_comp_DMFT}~(b), we show the self-energy at the antinode and at the node after the substitution of the local part of the self-energy (equation \ref{eq:sub_sigma_loc_dmft}). This substitution corrects the high-frequency behavior of the TPSC self-energy, while preserving the nodal-antinodal anisotropy at low frequency. As this substitution was found to yield accurate results in the one-band Hubbard model \cite{Martin_2023}, at least in the limit of validity of the TPSC approach, this method should also be an adequate way of including nonlocal fluctuations to DMFT for the Emery model.

% Conclusion
\section{Conclusion}

In this paper, we first used the functional derivative approach to obtain two equations for the self-energy in the Emery model. We then used this formalism to obtain the first level of approximation for the self-energy in the TPSC approach. This work lead us to the conclusion that, in models where the orbitals are not degenerate, using the noninteracting orbital densities for the TPSC calculation might lead to incorrect results.

We then applied the TPSC+DMFT approach to the Emery model, with the additional input of the interacting density from DMFT to the TPSC calculation. We considered a specific realization of the Emery model, where we showed that the noninteracting densities cannot be used reliably within TPSC. This is shown in Figures \ref{fig:docc_usp} and \ref{fig:selfEnergy_comp_DMFT}, where we compare the double occupancy and the local self-energy obtained from DMFT to our TPSC results. The TPSC calculations performed with the interacting orbital densities yield the most accurate results. In the spirit of previous work on the one-band Hubbard model \cite{Martin_2023}, we demonstrate that the TPSC approach can be used to include the effect of nonlocal fluctuations to DMFT in a multiorbital setting. In particular, we found the important result that the vertex for spin fluctuations $U^{\mathrm{sp}}_d$ decreases with filling, with important implications for electron-doped cuprates. 

Future work using this method could include the investigation of the antiferromagnetic pseudogap in the Emery model. 
Notably, it would be of interest to investigate whether the Vilk criterion formulated for the one-band Hubbard model also applies to the Emery model \cite{Vilk_1997,Kyung_2004} and whether it is possible to explain ARPES experiments without the need for ad-hoc sreening of the bare interaction. 
A host of new questions for TPSC also arise from this model such as orbital fluctuations. Our work also sets the stage for multiorbital calculations for pnictides, for strontium ruthenate and a host of multiband correlated electron systems where spin fluctuations dominate the physics.

\textit{Acknowledgments.} We are grateful to %Dominik Lessnich and 
David Sénéchal for useful discussions and to Yury Vilk for comments on the manuscript. This work has been supported by the Natural Sciences and Engineering Research Council of Canada (NSERC) under grant RGPIN-2019-05312, by a Vanier Scholarship from NSERC (C. G.-N.), by an Undergraduate Student Research Award from NSERC (J. L.), by an Excellence Scolarship from Hydro-Qu\'ebec (N. M.) and by the Canada First Research Excellence Fund. Simulations were performed on computers provided by the Canadian Foundation for Innovation, the Minist\`ere de l'\'Education des Loisirs et du Sport (Qu\'ebec), Calcul Qu\'ebec, and the Digital Research Alliance of Canada.  

% Annexes
\appendix
\section{Calculation details: source field approach for a simple multiorbital hamiltonian}
\label{app:details_source_fields}
We first obtain the equations of motion for $c_{l\gamma\sigma}(\tau)$ by computing its imaginary time derivative
%%%%%%%%%%%%%%%%
\begin{equation}
    \frac{\partial c_{l\gamma\sigma'}(\tau)}{\partial \tau} = \left [ H-\mu\sum_{i\alpha}n_{i\alpha}, c_{l\gamma\sigma'}(\tau) \right ].
\end{equation}
%%%%%%%%%%%%%%%%
To do so, we use the identity
%%%%%%%%%%%%%%%%
\begin{align}
    [AB,C] &=  A\{B,C\} - \{A,C\} B,\\
    \Rightarrow[c^\dagger_{i\alpha\sigma}c_{j\beta\sigma},c_{l\gamma\sigma'}] &= - \delta_{il}\delta_{\alpha\gamma}\delta_{\sigma\sigma'}c_{j\beta\sigma}.
\end{align}
%%%%%%%%%%%%%%%%
We evaluate the commutator with each term of the hamiltonian and obtain
%%%%%%%%%%%%%%%%
\begin{equation}
    \sum_{ij}\sum_{\alpha\beta}\sum_{\sigma}t_{ij,\alpha\beta}[c^\dagger_{i\alpha\sigma}c_{j\beta\sigma}, c_{l\gamma\sigma'}] = -\sum_{j\beta}t_{lj,\gamma\beta}c_{j\beta\sigma'},
    \label{eq:Comm_Kin}
\end{equation}
%%%%%%%%%%%%%%%%
%%%%%%%%%%%%%%%%
\begin{equation}
    -\mu\sum_{i}\sum_{\alpha}\sum_\sigma [c^\dagger_{i\alpha\sigma}c_{i\alpha\sigma}, c_{l\gamma\sigma'}] = \mu c_{l\gamma\sigma'},
    \label{eq:Comm_mu}
\end{equation}
%%%%%%%%%%%%%%%%
%%%%%%%%%%%%%%%%
\begin{equation}
    \sum_{i}\sum_{\alpha\beta}U_{\alpha\beta}[n_{i\alpha\uparrow}n_{i\beta\downarrow},c_{l\gamma\sigma'}] = -\sum_{\alpha}U_{\alpha\gamma}n_{l\alpha-\sigma'}c_{l\gamma\sigma'}.
    \label{eq:Comm_U}
\end{equation}
%%%%%%%%%%%%%%%%
Hence, the final result is
%%%%%%%%%%%%%%%%
\begin{align}
    \frac{\partial c_{l\gamma\sigma'}(\tau)}{\partial \tau} = &-t_{l\bar{j},\bar{\beta}\gamma}c_{\bar{j}\bar{\beta}\sigma'}(\tau)+\mu c_{l\gamma\sigma'}(\tau) \nonumber \\
    &-U_{\bar{\beta}\gamma}n_{l\bar{\beta}-\sigma'}(\tau)c_{l\gamma\sigma'}(\tau).
\end{align}
%%%%%%%%%%%%%%%%

To obtain the adjoint equation for the self-energy, we need
%%%%%%%%%%%%%%%%
\begin{equation}
    \frac{\partial c^\dagger_{l\gamma\sigma'}(\tau)}{\partial \tau} = \left [ H-\mu\sum_{i\alpha}n_{i\alpha}, c^\dagger_{l\gamma\sigma'}(\tau) \right ].
\end{equation}
%%%%%%%%%%%%%%%%
%To do so, we use the identity
%%%%%%%%%%%%%%%%
\begin{align}
%    [AB,C] &\Rightarrow A\{B,C\} - \{A,C\} B,\\
 [c^\dagger_{i\alpha\sigma}c_{j\beta\sigma},c^\dagger_{l\gamma\sigma'}] &= \delta_{jl}\delta_{\beta\gamma}\delta_{\sigma\sigma'}c^\dagger_{i\alpha\sigma}.
\end{align}
%%%%%%%%%%%%%%%%
We evaluate the commutator with each term of the hamiltonian and obtain
%%%%%%%%%%%%%%%%
\begin{equation}
%    \sum_{ij}\sum_{\alpha\beta}\sum_{\sigma}t_{ij,\alpha\beta}[c^\dagger_{i\alpha\sigma}c^\dagger_{j\beta\sigma}, c_{l\gamma\sigma'}] = \sum_{j\beta}t_{li,\gamma\beta}c^\dagger_{j\beta\sigma'},
    \sum_{ij}\sum_{\alpha\beta}\sum_{\sigma}t_{ij,\alpha\beta}[c^\dagger_{i\alpha\sigma}c_{j\beta\sigma}, c^\dagger_{l\gamma\sigma'}] = \sum_{i\alpha}t_{il,\alpha\gamma}c^\dagger_{i\alpha\sigma'}
    \label{eq:Comm_Kin_dagger}
\end{equation}
%%%%%%%%%%%%%%%%
%%%%%%%%%%%%%%%%
\begin{equation}
    -\mu\sum_{i}\sum_{\alpha}\sum_\sigma [c^\dagger_{i\alpha\sigma}c_{i\alpha\sigma}, c^\dagger_{l\gamma\sigma'}] = -\mu c^\dagger_{l\gamma\sigma'},
    \label{eq:Comm_mu_dagger}
\end{equation}
%%%%%%%%%%%%%%%%
%%%%%%%%%%%%%%%%
\begin{equation}
    \sum_{i}\sum_{\alpha\beta}U_{\alpha\beta}[n_{i\alpha\uparrow}n_{i\beta\downarrow},c^\dagger_{l\gamma\sigma'}] = \sum_{\alpha}U_{\alpha\gamma}n_{l\alpha-\sigma'}c^\dagger_{l\gamma\sigma'}.
    \label{eq:Comm_U_dagger}
\end{equation}
%%%%%%%%%%%%%%%%
%%%%%%%%%%%%%%%%
Hence, the final result is
%%%%%%%%%%%%%%%%
%%%%%%%%%%%%%%%%
\begin{align}
    \frac{\partial c^\dagger_{l\gamma\sigma'}(\tau)}{\partial \tau} = &t_{l\bar{j},\bar{\beta}\gamma}c^\dagger_{\bar{j}\bar{\beta}\sigma'}(\tau)-\mu c^\dagger_{l\gamma\sigma'}(\tau)\nonumber \\
    &+U_{\bar{\beta}\gamma}n_{l\bar{\beta}-\sigma'}(\tau)c^\dagger_{l\gamma\sigma'}(\tau).
\end{align}
%%%%%%%%%%%%%%%%

Let us recall the Green's function in the presence of a source field $\phi$. First, the partition function is 
%%%%%%%%%%%%%%%%
\begin{equation}
    Z[\phi] = \left \langle T_\tau \exp\left[-c^\dagger_{\bar{\alpha}\bar{\sigma}}(\bar{1})\phi_{\bar{\alpha}\bar{\beta}\bar{\sigma}}(\bar{1},\bar{2})c_{\bar{\beta}\bar{\sigma}}(\bar{2})\right]\right \rangle,
\end{equation}
%%%%%%%%%%%%%%%%
which leads to the definiton 
\begin{equation}
    S[\phi] = \exp\left[-c^\dagger_{\bar{\alpha}\bar{\sigma}}(\bar{1})\phi_{\bar{\alpha}\bar{\beta}\bar{\sigma}}(\bar{1},\bar{2})c_{\bar{\beta}\bar{\sigma}}(\bar{2})\right].
\end{equation}
%%%%%%%%%%%%%%%%
Then, the Green's function in the presence of the source field $\phi$ is defined as
%%%%%%%%%%%%%%%%
\begin{align}
    \mathcal{G}_{\alpha\beta\sigma}(1,2)_\phi&=-\frac{\delta \ln{(Z[\phi])}}{\delta \phi_{\beta \alpha \sigma}(2,1)},\\
    &= -\left \langle T_\tau c_{\alpha \sigma}(1)c^\dagger_{\beta\sigma}(2)\right \rangle_\phi,
\end{align}
%%%%%%%%%%%%%%%%
where the average of an operator $O(\tau_1,\dots,\tau_N)$ in the presence of the source field is defined by
%%%%%%%%%%%%%%%%
\begin{equation}
    \left \langle T_\tau O(\tau_1,\dots,\tau_N)\right \rangle_\phi \equiv \frac{1}{Z[\phi]}\left \langle T_\tau O(\tau_1,\dots,\tau_N)S[\phi]\right \rangle.
\end{equation}
%%%%%%%%%%%%%%%%

We are ready to obtain the equations of motion for the Green's function using the results for the commutators  ~\eref{eq:Comm_Kin} to \eref{eq:Comm_U}
%%%%%%%%%%%%%%%%
\begin{align}
    &\frac{\partial \mathcal{G}_{\alpha\beta\sigma}(1,2)_\phi}{\partial \tau_1} =  -t_{x_1\bar{j},\alpha\bar{\gamma}}\mathcal{G}_{\bar{\gamma}\beta\sigma}(\bar{x_j},\tau_1,2)_\phi + \mu\mathcal{G}_{\alpha\beta\sigma}(1,2)_\phi \nonumber \\
    &+ U_{\alpha\bar{\gamma}} \left \langle T_\tau c^\dagger_{\bar{\gamma}-\sigma}(1^+)c_{\bar{\gamma}-\sigma}(1)c_{\alpha\sigma}(1)c^\dagger_{\beta\sigma}(2)\right \rangle_\phi \nonumber \\
    &-\delta_{\alpha\beta}\delta(1-2)-\phi_{\alpha\bar{\gamma}\sigma}(1,\bar{3})\mathcal{G}_{\bar{\gamma}\beta\sigma}(\bar{3},2)_\phi.
    \label{eq:A_Dyson1_with_phi}
\end{align}
%%%%%%%%%%%%%%%%
The delta function comes from differentiating the time-ordering. The term involving $\phi$ is standard and comes from the derivative of the integral bounds when applying the time ordering operator. We rearrange the terms in order to define the noninteracting Green's function $\mathcal{G}^{(0)}$
%%%%%%%%%%%%%%%%
\begin{align}
    &\left [ \left ( \frac{\partial}{\partial \tau_1} -\mu \right ) \delta(\bar{x_j}-x_1)\delta_{\alpha\bar{\gamma}} + t_{x_1\bar{j},\alpha\bar{\gamma}}\right ]\mathcal{G}_{\bar{\gamma}\beta\sigma}(\bar{x_j},\tau_1,2)_\phi \nonumber \\
    &= -\delta_{\alpha\beta}\delta(1-2)-\phi_{\alpha\bar{\gamma}\sigma}(1,\bar{3})\mathcal{G}_{\bar{\gamma}\beta\sigma}(\bar{3},2)_\phi\nonumber \\
    &+ U_{\alpha\bar{\gamma}}\left \langle T_\tau c^\dagger_{\bar{\gamma}-\sigma}(1^+)c_{\bar{\gamma}-\sigma}(1)c_{\alpha\sigma}(1)c^\dagger_{\beta\sigma}(2)\right \rangle_\phi,\\
   &\equiv - \mathcal{G}^{(0)}_{\alpha\bar{\gamma}\sigma}(1,\bar{3})^{-1}\mathcal{G}_{\bar{\gamma}\beta\sigma}(\bar{3},2)_\phi.
\end{align}
%%%%%%%%%%%%%%%%
From Dyson's equation,
%%%%%%%%%%%%%%%%
\begin{align}
    \mathcal{G}_{\alpha\beta\sigma}(1,2)_\phi^{-1} = &\mathcal{G}^{(0)}_{\alpha\beta\sigma}(1,2)^{-1}\nonumber \\
    &-\phi_{\alpha\beta\sigma}(1,2)-\Sigma_{\alpha\beta\sigma}(1,2)_\phi,
    \label{eq:A_Dyson2_with_phi}
\end{align}
%%%%%%%%%%%%%%%%
we define the self-energy as
%%%%%%%%%%%%%%%%
\begin{align}
    \Sigma_{\alpha\bar{\gamma}\sigma}&(1,\bar{3})_\phi \mathcal{G}_{\bar{\gamma}\beta\sigma}(\bar{3},2)_\phi = \nonumber \\
    &-U_{\alpha\bar{\gamma}}\left \langle T_\tau c^\dagger_{\bar{\gamma}-\sigma}(1^+)c_{\bar{\gamma}-\sigma}(1)c_{\alpha\sigma}(1)c^\dagger_{\beta\sigma}(2)\right \rangle_\phi,
    \label{eq:app_selfEnergy_gen}
\end{align}
in agreement with \eref{eq:selfEnergy_gen}.
%%%%%%%%%%%%%%%%

The analogous approach to find the adjoint equation for the self-energy starts from the derivative with respect to $\tau_2$ instead of $\tau_1$. We obtain
%%%%%%%%%%%%%%%%
\begin{align}
    &\frac{\partial \mathcal{G}_{\alpha\beta\sigma}(1,2)_\phi}{\partial \tau_2} =  t_{\bar{j}x_2\bar{\gamma}\beta}\mathcal{G}_{\alpha\bar{\gamma}\sigma}(1,\bar{x_j},\tau_2)_\phi - \mu\mathcal{G}_{\alpha\beta\sigma}(1,2)_\phi \nonumber \\
    &- U_{\bar{\gamma}\beta} \left \langle T_\tau c^\dagger_{\bar{\gamma}-\sigma}(2^+)c_{\bar{\gamma}-\sigma}(2)c_{\alpha\sigma}(1)c^\dagger_{\beta\sigma}(2)\right \rangle \nonumber \\
    &+\delta_{\alpha\beta}\delta(1-2)+\mathcal{G}_{\alpha\bar{\gamma}\sigma}(1,\bar{3})_\phi \phi_{\bar{\gamma}\beta\sigma}(\bar{3},2).
\end{align}
%%%%%%%%%%%%%%%%
We rearrange the terms in order to define the noninteracting Green's function $\mathcal{G}^{(0)}$
%%%%%%%%%%%%%%%%
\begin{align}
    &\left [ \left (\frac{\partial}{\partial \tau_2}+\mu\right )\delta(\bar{x_j}-x_2)\delta_{\beta\bar{\gamma}} - t_{\bar{j}x_2\alpha\bar{\gamma}}\right ]\mathcal{G}_{\alpha\bar{\gamma}\sigma}(1,\bar{x_j},\tau_2)_\phi \nonumber \\
    &= \delta_{\alpha\beta}\delta(1-2)+\mathcal{G}_{\alpha\bar{\gamma}\sigma}(1,\bar{3})_\phi \phi_{\bar{\gamma}\beta\sigma}(\bar{3},2)\nonumber \\
    &- U_{\bar{\gamma}\beta}\left \langle T_\tau c^\dagger_{\bar{\gamma}-\sigma}(2^+)c_{\bar{\gamma}-\sigma}(2)c_{\alpha\sigma}(1)c^\dagger_{\beta\sigma}(2)\right \rangle,\\
   &\equiv \mathcal{G}_{\alpha\bar{\gamma}\sigma}(1,\bar{3})_\phi\mathcal{G}^{(0)}_{\bar{\gamma}\beta\sigma}(\bar{3},2)^{-1}
\end{align}
%%%%%%%%%%%%%%%%
From Dyson's equation,
%%%%%%%%%%%%%%%%
\begin{align}
    \mathcal{G}_{\alpha\beta\sigma}(1,2)_\phi^{-1} = &\mathcal{G}^{(0)}_{\alpha\beta\sigma}(1,2)^{-1}\nonumber \\
    &-\phi_{\alpha\beta\sigma}(1,2)-\Sigma_{\alpha\beta\sigma}(1,2)_\phi,
\end{align}
%%%%%%%%%%%%%%%%
we obtain a second definition for the self-energy
%%%%%%%%%%%%%%%%
\begin{align}
    \mathcal{G}_{\alpha \bar{\gamma}\sigma}&(1,\bar{3})_\phi \Sigma_{\bar{\gamma}\beta\sigma}(\bar{3},2)_\phi =\nonumber \\
    &-U_{\bar{\gamma}\beta}\left \langle T_\tau c^\dagger_{\bar{\gamma}-\sigma}(2^+)c_{\bar{\gamma}-\sigma}(2)c_{\alpha\sigma}(1)c^\dagger_{\beta\sigma}(2)\right \rangle,
    \label{eq:app_selfEnergy_gen_tau2}
\end{align}
in agreement with  \eref{eq:selfEnergy_gen_tau2}.
%%%%%%%%%%%%%%%%

\section{Effective one-band problem}
\label{app:effective_one_band}
Since only the $d$ orbitals are correlated in the Emery model, it is possible to find an effective one-band model. To see this, we first write the Dyson equation in matrix form, dropping spin indices
%%%%%%%%%%%%%%%%
\begin{equation}
    \mathbf{G}(k)^{-1} =  \mathbf{G}^{(0)}(k)^{-1} - \mathbf{\Sigma}(k),
\end{equation}
%%%%%%%%%%%%%%%%
where we define the matrices as
%%%%%%%%%%%%%%%%
\begin{equation}
    \mathbf{G}^{(0)}(k)^{-1} = (ik_n+\mu)\mathbf{1}-\mathbf{h}^{(0)}(\mathbf{k}),
\end{equation}
%\begin{equation}
%    \mathbf{G}(k)^{-1} = \begin{pmatrix}
%        ik_n+\mu-h^{(0)}_{d}(\mathbf{k})-\Sigma_d(k) & -\mathbf{h}^{(0)}_{dp}(\mathbf{k})-\mathbf{\Sigma}_{dp}(k),\\
%        -\mathbf{h}^{(0)}_{pd}(\mathbf{k})-\mathbf{\Sigma}_{pd}(k) & ik_n+\mu-\mathbf{h}^{(0)}_{p}(\mathbf{k})-\mathbf{\Sigma}_{p}(k) 
%    \end{pmatrix}.
%\end{equation}
%%%%%%%%%%%%%%%%
Using the following notation for matrix elements
%%%%%%%%%%%%%%%%
\begin{align}
    \mathbf{A} &= \begin{pmatrix}
        A_{dd} & A_{dp_x} & A_{dp_y}\\
        A_{p_xd} & A_{p_xp_x} & A_{p_xp_y}\\
        A_{p_yd} & A_{p_yp_x} & A_{p_yp_y}
    \end{pmatrix},\\
    &\equiv \begin{pmatrix}
        A_{d} & \mathbf{A}_{dp} \\
        \mathbf{A}_{pd} & \mathbf{A}_{p} 
    \end{pmatrix}.
\end{align}
%%%%%%%%%%%%%%%%
the self-energy takes the form
%%%%%%%%%%%%%%%%
\begin{equation}
    \mathbf{\Sigma}(k) = \begin{pmatrix}
        \Sigma_d(k) & \mathbf{\Sigma}_{dp}(k)\\
        \mathbf{\Sigma}_{pd}(k) & \mathbf{\Sigma}_{p}(k)
    \end{pmatrix}.
\end{equation}
%%%%%%%%%%%%%%%%

To rewrite this as an effective one-band problem, we use the identity
%%%%%%%%%%%%%%%%
\begin{equation}
   \mathbf{G}(k)^{-1}\mathbf{G}(k) = \mathbf{1}.
\end{equation}
%%%%%%%%%%%%%%%%
We also make use of the fact that, as seen in the previous subsection, the self-energy only has one non-zero matrix element,
%%%%%%%%%%%%%%%%
\begin{equation}
    \mathbf{\Sigma}(k) = \begin{pmatrix}
        \Sigma_d(k) & 0 & 0 \\
        0 & 0 & 0\\
        0 & 0 & 0
    \end{pmatrix}.
\end{equation}
%%%%%%%%%%%%%%%%
From the matrix multiplication, we obtain a system of linear equations from which we can find an expression for $\mathcal{G}_d$. More specifically, we use the equations
%%%%%%%%%%%%%%%%
\begin{align}
    \left (ik_n+\mu-h^{(0)}_d(\mathbf{k})-\Sigma_d(k)\right )\mathcal{G}_d(k)-\mathbf{h}^{(0)}_{dp}(\mathbf{k})\mathbf{G}_{pd}(k)=1,\\
    -\mathbf{h}^{(0)}_{pd}(\mathbf{k})\mathcal{G}_d(k)+\left (ik_n+\mu-\mathbf{h}^{(0)}_p(\mathbf{k})\right )\mathbf{G}_{pd}(k)=0.
\end{align}
%%%%%%%%%%%%%%%%
Subtituting $\mathbf{G}_{pd}(k)$ from the last equation in the previous one, we find
%%%%%%%%%%%%%%%%
\begin{widetext}
\begin{equation}
    \mathcal{G}_d(k) = \left [ ik_n+\mu-h^{(0)}_d(\mathbf{k})-\Sigma_d(k)-\mathbf{h}^{(0)}_{dp}(\mathbf{k})\left (ik_n+\mu-\mathbf{h}_p^{(0)}(\mathbf{k})\right )^{-1}\mathbf{h}^{(0)}_{pd}(\mathbf{k})\right ]^{-1}.
\end{equation}
\end{widetext}
%%%%%%%%%%%%%%%%
Hence, the Green's function for the $d$ orbitals contains the self-energy for the $d$ orbitals as well as a hybridization function, $\Delta_{dp}$ defined as
%%%%%%%%%%%%%%%%
\begin{equation}
    \Delta_{dp}(\mathbf{k},ik_n) \equiv \mathbf{h}^{(0)}_{dp}(\mathbf{k})\left (ik_n+\mu-\mathbf{h}_p^{(0)}(\mathbf{k})\right )^{-1}\mathbf{h}^{(0)}_{pd}(\mathbf{k}). 
\end{equation}
%%%%%%%%%%%%%%%%

%\bibliography{bibli.bib}


%%%%%%%%%% Merge with supplemental materials %%%%%%%%%%
%%%%%%%%%% Prefix a "S" to all equations, figures, tables and reset the counter %%%%%%%%%%

% \setcounter{equation}{0}
% \setcounter{figure}{0}
% \setcounter{table}{0}
% \setcounter{page}{1}
% \makeatletter

% \renewcommand{\theequation}{S\arabic{equation}}
% \renewcommand{\thefigure}{S\arabic{figure}}
% \renewcommand{\thetable}{S\arabic{table}}
% \renewcommand{\bibnumfmt}[1]{[S#1]}
% \renewcommand{\citenumfont}[1]{S#1}
% %%%%%%%%%% Prefix a "S" to all equations, figures, tables and reset the counter %%%%%%%%%%

% \newpage 

% \newpage

% \begin{appendix}

%     \begin{center}
%         \textbf{\large SUPPLEMENTAL MATERIAL}
%         \linebreak
    
%         \textbf{\large An Improved Two-Particle Self-Consistent Approach}
%         \linebreak
    
%         C. Lahaie, Y. Wang, C. Gauvin-Ndiaye, Y. M. Vilk, and A.-M. S. Tremblay
%     \end{center}

%     In this Supplemental Material, we discuss in turn
%     \begin{itemize}
%         \item A thing,
%         \item And another thing.
%     \end{itemize}

%     \subsection{A thing}
    
%     \subsection{And another thing}

% \end{appendix}

\end{document}